\documentclass{iopjournal}

\usepackage{amsmath,amsfonts}
\usepackage{dsfont}
\usepackage{bm}
\usepackage{color}
\usepackage{comment}
\usepackage{graphicx}

\usepackage{subcaption}
\usepackage{hyperref}
\usepackage{tikz}
\usepackage{pgfplots}
\pgfplotsset{compat=1.6}

\usetikzlibrary{spy}
\usepackage{placeins}

\newcommand{\hyp}[4]{\, {}_{2}F_{1} \left( \begin{matrix} #1,\  #2 \\ #3  \end{matrix} \, \Bigg \vert \, #4 \right )} 

\newcommand{\intl}{\int \limits_{0}^{2 \pi} \mathrm{d}}

\newcommand{\intla}[2]{\int \limits_{#1}^{#2} \mathrm{d}}

\newcommand{\WsqAna}{
(1, 4.0) (3, 5.875) (5, 7.4008) (7, 8.6679) (9, 9.7725) (11, 10.764) (13, 11.672) (15, 12.515) (17, 13.304) (19, 14.048) (21, 14.756) (23, 15.43) (25, 16.077) (27, 16.699) (29, 17.298) (31, 17.877) (33, 18.438) (35, 18.982) (37, 19.512) (39, 20.027) (41, 20.529) (43, 21.019) (45, 21.499) (47, 21.967) (49, 22.426) (51, 22.876) (53, 23.317) (55, 23.75) (57, 24.175) (59, 24.592) (61, 25.003) (63, 25.407) (65, 25.805) (67, 26.197) (69, 26.583) (71, 26.963) (73, 27.338) (75, 27.709) (77, 28.074) (79, 28.434) (81, 28.79) (83, 29.142) (85, 29.489) (87, 29.833) (89, 30.172) (91, 30.508) (93, 30.84) (95, 31.169) (97, 31.494) (99, 31.816) (101, 32.134) (103, 32.449) (105, 32.762) (107, 33.071) (109, 33.378) (111, 33.682) (113, 33.983) (115, 34.281) (117, 34.577) (119, 34.87) (121, 35.161) (123, 35.45) (125, 35.736) (127, 36.02) (129, 36.302) (131, 36.581) (133, 36.858) (135, 37.134) (137, 37.407) (139, 37.678) (141, 37.948) (143, 38.215) (145, 38.481) (147, 38.745) (149, 39.007) (151, 39.267) (153, 39.526) (155, 39.782) (157, 40.038) (159, 40.291) (161, 40.543) (163, 40.794) (165, 41.043) (167, 41.29) (169, 41.536) (171, 41.781) (173, 42.024) (175, 42.265) (177, 42.506) (179, 42.745) (181, 42.982) (183, 43.219) (185, 43.454) (187, 43.687) (189, 43.92) (191, 44.151) (193, 44.381) (195, 44.61) (197, 44.838) (199, 45.065) (201, 45.29) (203, 45.514) (205, 45.738) (207, 45.96) (209, 46.181) (211, 46.401) (213, 46.62) (215, 46.838) (217, 47.055) (219, 47.271) (221, 47.486) (223, 47.7) (225, 47.913) (227, 48.125) (229, 48.336) (231, 48.547) (233, 48.756) (235, 48.964) (237, 49.172) (239, 49.379) (241, 49.585) (243, 49.79) (245, 49.994) (247, 50.197) (249, 50.4) (251, 50.601)
}

\newcommand{\WsqMC}{
(1,{8330132/2083125}) +- (0,0.056555) (10,{63509471/6249375}) +- (0,0.14373) (20,{9950896/694375}) +- (0,0.20268) (30,{4421024/249975}) +- (0,0.25013) (40,{14409831/694375}) +- (0,0.29350) (50,{141661351/6249375}) +- (0,0.32059) (60,{50806733/2083125}) +- (0,0.34494) (70,{18841079/694375}) +- (0,0.38375) (80,{5482607/189375}) +- (0,0.40945) (90,{186799744/6249375}) +- (0,0.42274) (100,{200969596/6249375}) +- (0,0.45481) (110,{208054816/6249375}) +- (0,0.47084) (120,{216621439/6249375}) +- (0,0.49023) (130,{9135431/249975}) +- (0,0.51686) (140,{237715904/6249375}) +- (0,0.53797) (150,{241299391/6249375}) +- (0,0.54608) (160,{84744448/2083125}) +- (0,0.57535) (170,{7928992/189375}) +- (0,0.59215) (180,{265275644/6249375}) +- (0,0.60034) (190,{109741/2525}) +- (0,0.61467) (200,{282286519/6249375}) +- (0,0.63884) (210,{295322039/6249375}) +- (0,0.66834) (220,{3900197/83325}) +- (0,0.66198) (230,{300697564/6249375}) +- (0,0.68050) (240,{314388911/6249375}) +- (0,0.71149) (250,{35729271/694375}) +- (0,0.72772)
}

\newcommand{\COneBulkMC}{
(0.1, -5.5068) +- (0, 0.1431) (0.2, -2.651) +- (0, 0.0823) (0.3, -1.6232) +- (0, 0.059) (0.4, -1.0067) +- (0, 0.0557) (0.5, -0.6554) +- (0, 0.0649) (0.6, -0.3416) +- (0, 0.0698) (0.7, -0.2227) +- (0, 0.0604) (0.8, 0.026) +- (0, 0.0744) (0.9, 0.2334) +- (0, 0.0588) (1.0, 0.4908) +- (0, 0.0583) (1.1, 0.7577) +- (0, 0.0565) (1.2, 1.2318) +- (0, 0.0695) (1.3, 1.8269) +- (0, 0.0773) (1.4, 3.4855) +- (0, 0.0799)
}

\newcommand{\COneSpecAna}{
(-2.0, 0.2587) (-1.9, 0.2734) (-1.8, 0.2901) (-1.7, 0.309) (-1.6, 0.3309) (-1.5, 0.3565) (-1.4, 0.3871) (-1.3, 0.4243) (-1.2, 0.4706) (-1.1, 0.529) (-1.0, 0.6027) (-0.9, 0.6935) (-0.8, 0.7999) (-0.7, 0.913) (-0.6, 1.0145) (-0.5, 1.0762) (-0.4, 1.0642) (-0.3, 0.9495) (-0.2, 0.7199) (-0.1, 0.3895) (0.0, 0.0) (0.1, -0.3895) (0.2, -0.7199) (0.3, -0.9495) (0.4, -1.0642) (0.5, -1.0762) (0.6, -1.0145) (0.7, -0.913) (0.8, -0.7999) (0.9, -0.6935) (1.0, -0.6027) (1.1, -0.529) (1.2, -0.4706) (1.3, -0.4243) (1.4, -0.3871) (1.5, -0.3565) (1.6, -0.3309) (1.7, -0.309) (1.8, -0.2901) (1.9, -0.2734) (2.0, -0.2587)
}

\newcommand{\COneSpecMC}{
(-1.9, 0.21) +- (0, 0.0107) (-1.7, 0.2787) +- (0, 0.0112) (-1.5, 0.315) +- (0, 0.0114) (-1.3, 0.402) +- (0, 0.0123) (-1.1, 0.5172) +- (0, 0.0112) (-0.9, 0.6533) +- (0, 0.0183) (-0.7, 0.9254) +- (0, 0.0152) (-0.5, 1.0635) +- (0, 0.0115) (-0.3, 0.9477) +- (0, 0.0078) (-0.2, 0.7318) +- (0, 0.0068) (-0.1, 0.3848) +- (0, 0.0047) (0.0, 0.0) +- (0, 0.0) (0.1, -0.3848) +- (0, 0.0047) (0.2, -0.7318) +- (0, 0.0068) (0.3, -0.9477) +- (0, 0.0078) (0.5, -1.0635) +- (0, 0.0115) (0.7, -0.9254) +- (0, 0.0152) (0.9, -0.6533) +- (0, 0.0183) (1.1, -0.5172) +- (0, 0.0112) (1.3, -0.402) +- (0, 0.0123) (1.5, -0.315) +- (0, 0.0114) (1.7, -0.2787) +- (0, 0.0112) (1.9, -0.21) +- (0, 0.0107)
}

\newcommand{\CTwoBulkAna}{
(-1.6, 0.3883) (-1.5, 0.4395) (-1.4, 0.5001) (-1.3, 0.5716) (-1.2, 0.6555) (-1.1, 0.753) (-1.0, 0.8647) (-0.9, 0.9902) (-0.8, 1.1281) (-0.7, 1.2749) (-0.6, 1.4257) (-0.5, 1.5739) (-0.4, 1.7116) (-0.3, 1.8303) (-0.2, 1.9221) (-0.1, 1.9801) (0.0, 2.0) (0.1, 1.9801) (0.2, 1.9221) (0.3, 1.8303) (0.4, 1.7116) (0.5, 1.5739) (0.6, 1.4257) (0.7, 1.2749) (0.8, 1.1281) (0.9, 0.9902) (1.0, 0.8647) (1.1, 0.753) (1.2, 0.6555) (1.3, 0.5716) (1.4, 0.5001) (1.5, 0.4395) (1.6, 0.3883)
}

\newcommand{\CTwoBulkMC}{
(-1.6, 0.325) +- (0, 0.0615) (-1.4, 0.4835) +- (0, 0.0712) (-1.2, 0.8082) +- (0, 0.1072) (-1.0, 0.9294) +- (0, 0.0549) (-0.8, 1.193) +- (0, 0.0564) (-0.6, 1.4435) +- (0, 0.0527) (-0.4, 1.73) +- (0, 0.0677) (-0.2, 2.0473) +- (0, 0.0952) (-0.05, 2.1584) +- (0, 0.361) (0.2, 1.9872) +- (0, 0.0985) (0.4, 1.7341) +- (0, 0.0589) (0.6, 1.5208) +- (0, 0.0547) (0.8, 1.1993) +- (0, 0.0486) (1.0, 0.9221) +- (0, 0.0572) (1.2, 0.6209) +- (0, 0.0693) (1.4, 0.4832) +- (0, 0.0657) (1.6, 0.3541) +- (0, 0.0781)
}

\newcommand{\CTwoSpecMC}{
(-1.5, 0.2319) +- (0, 0.0888) (-1.4, 0.3766) +- (0, 0.077) (-1.3, 0.4558) +- (0, 0.0768) (-1.2, 0.693) +- (0, 0.0708) (-1.1, 1.0166) +- (0, 0.0727) (-1.0, 1.4535) +- (0, 0.0657) (-0.9, 2.1506) +- (0, 0.0802) (-0.8, 3.0569) +- (0, 0.1122) (-0.7, 4.8141) +- (0, 0.1581) (-0.6, 7.6879) +- (0, 0.3059) (-0.4, 6.3548) +- (0, 0.2597) (-0.3, 3.6189) +- (0, 0.1143) (-0.2, 2.3846) +- (0, 0.0601) (-0.1, 1.9973) +- (0, 0.0358) (0.0, 2.0581) +- (0, 0.0249) (0.1, 2.343) +- (0, 0.0356) (0.2, 2.6944) +- (0, 0.06) (0.3, 2.9529) +- (0, 0.1145) (0.4, 3.2315) +- (0, 0.2603) (0.6, 2.9517) +- (0, 0.3067) (0.7, 2.7949) +- (0, 0.1586) (0.8, 2.664) +- (0, 0.1123) (0.9, 2.276) +- (0, 0.0802) (1.0, 1.9633) +- (0, 0.0656) (1.1, 1.6512) +- (0, 0.0726) (1.2, 1.418) +- (0, 0.0707) (1.3, 1.2222) +- (0, 0.0767) (1.4, 0.998) +- (0, 0.0769) (1.5, 0.8565) +- (0, 0.0888)
}

\newcommand{\CTwoSpecAna}{
(-1.5, 0.3957) (-1.475, 0.4238) (-1.45, 0.455) (-1.425, 0.4897) (-1.4, 0.5282) (-1.375, 0.5709) (-1.35, 0.6182) (-1.325, 0.6706) (-1.3, 0.7286) (-1.275, 0.7928) (-1.25, 0.8636) (-1.225, 0.9417) (-1.2, 1.0277) (-1.175, 1.1223) (-1.15, 1.2263) (-1.125, 1.3403) (-1.1, 1.4653) (-1.075, 1.602) (-1.05, 1.7514) (-1.025, 1.9145) (-1.0, 2.0923) (-0.975, 2.286) (-0.95, 2.4967) (-0.925, 2.7258) (-0.9, 2.9749) (-0.875, 3.2456) (-0.85, 3.54) (-0.825, 3.8604) (-0.8, 4.2097) (-0.775, 4.5915) (-0.75, 5.0104) (-0.725, 5.4725) (-0.7, 5.9865) (-0.675, 6.5643) (-0.65, 7.2243) (-0.625, 7.9952) (-0.6, 8.9269) (nan, nan)

 (-0.4, 7.6337) (-0.375, 6.5973) (-0.35, 5.7648) (-0.325, 5.0806) (-0.3, 4.5112) (-0.275, 4.0349) (-0.25, 3.6363) (-0.225, 3.3045) (-0.2, 3.0307) (-0.175, 2.8079) (-0.15, 2.6304) (-0.125, 2.4931) (-0.1, 2.3915) (-0.075, 2.3215) (-0.05, 2.2792) (-0.025, 2.2613) (0.0, 2.2643) (0.025, 2.285) (0.05, 2.3206) (0.075, 2.3682) (0.1, 2.4252) (0.125, 2.4891) (0.15, 2.5575) (0.175, 2.6285) (0.2, 2.7) (0.225, 2.7704) (0.25, 2.838) (0.275, 2.9015) (0.3, 2.9598) (0.325, 3.012) (0.35, 3.0571) (0.375, 3.0946) (0.4, 3.1241) (0.425, 3.1453) (0.45, 3.158) (0.475, 3.1622) (0.5, 3.1581) (0.525, 3.1459) (0.55, 3.1259) (0.575, 3.0985) (0.6, 3.0641) (0.625, 3.0233) (0.65, 2.9766) (0.675, 2.9246) (0.7, 2.8678) (0.725, 2.8069) (0.75, 2.7424) (0.775, 2.6749) (0.8, 2.6049) (0.825, 2.5331) (0.85, 2.4598) (0.875, 2.3855) (0.9, 2.3108) (0.925, 2.2359) (0.95, 2.1613) (0.975, 2.0872) (1.0, 2.0139) (1.025, 1.9418) (1.05, 1.8709) (1.075, 1.8015) (1.1, 1.7338) (1.125, 1.6678) (1.15, 1.6036) (1.175, 1.5414) (1.2, 1.4812) (1.225, 1.423) (1.25, 1.3669) (1.275, 1.3128) (1.3, 1.2607) (1.325, 1.2107) (1.35, 1.1626) (1.375, 1.1165) (1.4, 1.0723) (1.425, 1.0299) (1.45, 0.9894) (1.475, 0.9507) (1.5, 0.9136)
}

\begin{document}

\articletype{Paper} 

\title{Winding Number Statistics of a Parametric Chiral Symplectic Random Matrix Ensemble}

\author{Pascal Toschka$^{1*}$, Luke Haas$^{1*}$, Thomas Guhr$^{1}$, Omri Gat$^{2}$ and Mario Kieburg$^{3}$}

\affil{$^1$ Fakult\"at f\"ur Physik, Universit\"at Duisburg--Essen, Lotharstr. 1, Duisburg 47057, Germany}

\affil{$^2$ The Racah Institute of Physics, The Hebrew University of Jerusalem, Levin Building, Sderot Magnes, Jerusalem 919040, Israel}

\affil{$^3$ School of Mathematics and Statistics, University of Melbourne, 813 Swanston Street, Parkville, Melbourne VIC 3010, Australia}

\affil{$^*$ Both authors contributed equally.}

\email{pascaltoschka@gmail.com, luke.haas@icloud.com, thomas.guhr@uni-due.de, omrigat@mail.huji.ac.il, m.kieburg@unimelb.edu.au}


\begin{abstract}
  The winding number is a simple topological invariant. In the case of
  chiral symmetry it characterises gapped phases of Fermions. We study
  statistical properties of this topological index or invariant in a
  chiral symplectic setting using Random Matrix Theory. We consider
  ensembles of Hamilton matrices in the symmetry class CII (quaternionic matrices) according
  to the classification in the tenfold way. We set up a parametric
  random matrix model and derive expressions for parametric
  correlations of the winding number density as well as for the
  discrete winding number distribution. We found
  a super-universality in the limit of large matrix dimensions for
  the one- and two-point correlators for the bulk of parameters, meaning that the results agree up to rescaling with those of the class AIII (complex matrices). In this context we employ a new
  method of unfolding and discover the Gaussian behaviour of the winding
  number distribution.
\end{abstract}

\section{Introduction}
\label{intro}

Gapped Hamiltonians appear frequently in quantum physics, particularly
in condensed matter, quantum chromodynamics (QCD) and quantum optics (e.g., in cavity quantum electrodynamics (QED) or atoms trapped in optical lattices). Statistical topology aims at
unravelling universal properties of such systems. To this end, as
little input as possible is employed, namely only the governing
invariances and symmetries as well as randomness.  The tenfold way~\cite{Altland1997,Chiu2016,Schnyder2008,Zirnbauer2021} classifies
systems according to the relevant invariances and symmetries: time
reversal invariance, particle-hole and chiral symmetries. The
bijection between the resulting classes and the Cartan symmetric
spaces prompts usage of the Cartan labels, see also
Ref.~\cite{Heinzner2005}.

We consider Random Matrix Theory (RMT) for general models of gapped
Hamiltonians or Dirac operators with a parametric dependence that does not
alter the system symmetries. In this parameter space, random curves or
manifolds are described whose different statistical properties are the
object of interest. In the context of chirality, RMT and the related
theory of Anderson localisation in mesoscopic disordered systems~\cite{Shapiro1993,Gade1993}, including the supersymmetry approach~\cite{Efetov1983,Efetov1996}, reveal
connections to QCD, especially to infrared level statistics and to the chiral phase transition. Chiral RMT was
introduced in Refs.~\cite{Shuryak1993,Verbaarschot1994,Verbaarschot2000}. The connection between Thouless energy and
Gell-Mann--Oakes--Renner relation was established in Refs.~\cite{Janik1998,Osborn1998} for a nonlinear $\sigma$ model with disorder and
chirality, see Ref.~\cite{Guhr2000}.

As always in RMT~\cite{Guhr1998,Mehta2004}, the limit of infinite
matrix dimension combined with unfolding is mandatory to arrive at
universal results independent of the probability density for the
random Hamiltonians~\cite{Braun2022}. Yet, another aspect of
universality enters since it has to be ensured that the choice of the
parameter dependence does not lead to nonuniversal features~\cite{Hahn2025}.  Here, we continue our study of the chiral
classes AIII, BDI and CII~\cite{Braun2022,Hahn2023b,Hahn2025} in which
time reversal invariance and particle hole symmetry together form the
chiral operator~\cite{Zirnbauer2021}.

In general, the possible types of topological invariants, which
manifest themselves as physical phases~\cite{Kitaev2009,Maffei2018},
depend on the dimension of the parameter space. We focus on a
one--dimensional parameter dependence and thus on the winding
number given by a closed loop in the random matrix space. The bulk--boundary correspondence implies that a non--zero
winding number corresponds to the exact number of edge states in an
open system~\cite{Prodan2016}. In the AIII case it was shown that
the winding number is a robust topological invariant
\cite{MondragonShem2014} even in the presence of strong disorder. The
Kitaev chain~\cite{Kitaev2009} is a prominent example for a system
with phases characterised by the winding number. Its topological phase
hosts Majorana zero and edge states, which can be experimentally
realised with superconducting heterostructures. They offer great
potential in topological quantum computing due to their non--Abelian
exchange statistics \cite{Alicea2012}. For an elementary introduction
to statistical topology and the Kitaev chain, see Ref.~\cite{Guhr2023}.

The present article is based on previous ideas~\cite{Braun2022,Hahn2023a,Hahn2023b,Hahn2025,Yahiaoui2025} where some of the authors introduced random matrix fields over the complex unit circle modelling a closed path in the Brillouin zone. In these works, we have mostly worked with Gaussian two-matrix models, though it was shown in~\cite{Yahiaoui2025} that general Gaussian random matrix fields and even non-Gaussian fields can be studied, as well, as long as the algebraic structures involved, such as determinants and Pfaffians, are still present. Explicit expressions for moment-generating functions of the winding number densities at finite matrix dimension have been derived in~\cite{Hahn2023a,Hahn2023b,Yahiaoui2025} for all three chiral symmetry classes AIII (complex aka unitary symmetry), CII (quaternion aka symplectic symmetry) and BDI (real aka orthogonal symmetry). However, only for the complex case we have derived the whole winding number statistics in~\cite{Braun2022,Hahn2025} as its determinantal expression has been the simplest.  We aim to derive expressions for the parametric correlations of the winding number density as well as for the winding number distribution and its first two moments for class CII in the limit of large matrix dimension. Furthermore, we explicitly address universality of these quantities in the limit of infinite matrix dimension where we find a Gaussian distribution of the rescaled winding number as it was observed for AIII-Hamiltonians~\cite{Braun2022,Hahn2025}.

The paper is organised as follows. In Sec.~\ref{sec:setup},
we briefly review the parametric random matrix model and define the winding
number and the winding number density. In Sec.~\ref{sec:finite.N} we introduce the statistical quantities and present
explicit results for finite $N$. The detailed derivations are given in Appendices~\ref{sec:kpoint} and \ref{sec:Derivations.D}. Section~\ref{sec:unfolding} details the unfolding of the winding number distribution, while Sec.~\ref{sec:unfold} derives the limit of large matrix dimension of the one- and two-point correlators, with the technical details relegated to Appendices~\ref{sec:rm} and \ref{sec:unfold.der} respectively. Finally, in Sec.~\ref{sec:conclusion} we summarise our findings.

\section{Posing the Problem}
\label{sec:setup}

We consider invertible chiral symplectic Hamiltonians $H(p)$ in CII as classified in the tenfold way~\cite{Altland1997,Chiu2016}, meaning $\det H(p)\neq0$. While the 1D topological phases with integer winding numbers are famously realised in classes BDI or AIII by the Kitaev chain or the Su–Schrieffer–Heeger model~\cite{SSH1979,Kitaev2001}, class CII is more difficult to access in condensed matter physics due to its negative-square particle-hole symmetry ($\mathcal{P}^2=-1$). The chiral symmetry is expressed in terms of the anticommutator
\begin{equation}
	\lbrace H(p), \mathcal{C} \rbrace =H(p)\mathcal{C}+\mathcal{C}H(p)= 0 
\end{equation}
with the unitary, idempotent chirality operator $\mathcal{C}$. In the framework of condensed matter physics $\mathcal{C} $ may be viewed as the composition of time reversal and the exchange of particle and hole states. The parameter $p\in[0,2\pi]$ is, then, the parametrisation of a wave vector in the Brillouin zone.

Following~\cite{Hahn2023a} in the case of $4N \times 4N$ matrices, $\mathcal{C}$ takes in its eigenbasis the form
\begin{equation}
	\mathcal{C} = \begin{bmatrix}
		\mathds{1}_{2N}& 0 \\
		0 & -\mathds{1}_{2N}  
	\end{bmatrix}
\end{equation}
implying an off-diagonal form of
\begin{equation}
	H(p) = \begin{bmatrix}
		0 & K(p) \\
		K^{\dagger}(p) & 0
	\end{bmatrix}    .
\end{equation}
The symplectic symmetry of the Hamiltonian---reflecting that it must lie in the class CII---is given by the anti-unitary symmetry
\begin{equation}
	\mathcal{T} K(p) \mathcal{T}^{-1}= K(-p), \quad \text{with }\quad \mathcal{T} = (i \sigma_2 \otimes \mathds{1}_N )\mathcal{K},
\end{equation}
where $\mathcal{T}$ is the time-reversal operator comprised of the complex conjugation $\mathcal{K}$ and the second Pauli matrix $\sigma_2 $. The time-reversal operator $\mathcal{T}$ effectively acts as a reflection inside the Brillouin zone by mapping $p $ onto $-p$ while performing a $\pi/2$ rotation in the spin space.

The Hamiltonian $H(p)=H(p+2\pi)$ as well as its off-diagonal block $K(p)=K(p+2\pi)$ are $2\pi$-periodic. This is, however, not necessarily the case for the eigenvectors of $K(p)$ nor its complex eigenvalues, although for the combined spectrum it is. The eigenvalues permute in general after one period, while the eigenvectors even acquire a complex phase. The invertibility of $H(p)$ carries over to $K(p)$. In particular, the determinant $\det K(p)\neq0$ is non-zero and $2\pi$-periodic as it is the product of all eigenvalues. Hence,  $\det K(p)$ can wind around the origin, which gives rise to the winding number
\begin{equation}
	W = \frac{1}{2 \pi i} \intl  p\, w(p)\qquad{\rm with}\quad w(p) = \frac{d}{dp} \ln \det K(p),\label{eq:definitionw}
\end{equation}
as a $2\mathbb{Z}$ topological index belonging to the symmetry class CII and spatial dimension one.
We call $w(p)$ the winding number density.

The setting considered in~\cite{Braun2022,Hahn2023a,Hahn2023b,Hahn2025} is the two matrix model
\begin{equation}
	K(p) = a(p)K_1 + b(p) K_2 = v^{T}(p) \begin{bmatrix}
		K_1 \\ K_2
	\end{bmatrix}.  \label{eq:modeloriginal}
\end{equation}
For the class CII, both $2N \times 2N$ matrices $K_1$ and $K_2$ are real quaternion Ginibre matrices~\cite{Ginibre1965}, which are related to the chiral Gaussian symplectic ensemble (chGSE). For simplicity, we choose $v(p) =  ( \cos p, i \sin p)$.
The parametrisation carries over to the Bloch Hamiltonian
\begin{equation}
	H(p) = \cos p\,H_1  + i \sin p\,H_2  \qquad \text{with} \quad H_j = \begin{bmatrix}
		0 & K_j \\
		K_j^{\dagger} & 0
	\end{bmatrix}.
\end{equation}
Exploring more general choices of $v(p)$ would allow one to probe more detailed features of the underlying eigenvalue statistics.

\section{Explicit Results at Finite Matrix dimension}\label{sec:finite.N}

The original method also pursued in~\cite{Braun2022,Hahn2023a} is via the $k$-point correlators $C_k(p)$ at $k$ different points $p_1,\ldots, p_k$ on the unit circle via the ensemble average
\begin{equation}
	C_k(p) = C_k(p_1,\dots,p_k) = \langle w(p_1) \dots w(p_k) \rangle , \label{eq:definitionck}
\end{equation}
meaning the $k$th moment of the winding number $W$ is
\begin{equation}\label{eq:momente}
\left\langle W^k\right\rangle=\left(\prod\limits_{j=1}^k\frac{1}{2 \pi i} \intl  p_j\right)C_k(p_1,\dots,p_k)
\end{equation}
The $k$-point correlators are generated by the $k$-fold derivative of an expectation value of ratios of determinants, i.e.,
\begin{equation}
	C_{k}(p) = \prod\limits_{n = 1}^k\frac{\partial }{ \partial p_{n}} Z_{k|k}(p,q) \Bigg \vert_{p = q} \qquad{\rm with}\quad Z_{k|k}(p,q) = \left \langle \frac{\prod_{n=1}^{k} \det K(p_n)}{\prod_{m=1}^{k} \det K(q_n)}\right\rangle. \label{eq:ckconstruct}
\end{equation}
The generating function $Z_{k|k}(p,q)$ has been calculated for all corresponding chiral classes AIII, CII and BDI in \cite{Hahn2023a,Hahn2023b} for a  two-matrix model with a general choice of $v(p)$ at finite matrix dimension $N$.    

In Sec.~\ref{sec:correlators} we will recall the results of~\cite{Hahn2023a}. Furthermore, we carry out the derivatives in~\eqref{eq:ckconstruct} and write explicitly what the one-point ($k=1$) and two-point ($k=2$) correlators are. Explicit results for the first and second moment at finite matrix dimension are derived in Sec.~\ref{sec:secw1}. Therein, we will also give the asymptotic scaling of the variance which determines the unfolding scaling in the ensuing section.

\subsection{Parametric correlations of the winding number density}
\label{sec:correlators}

In the chiral symplectic case, the generator for equal number $k$ of determinants in numerator and denominator has been found to be \cite{Hahn2023a} 
\begin{align}
	Z_{k|k}(p,q) = \frac{1 }{\det \begin{bmatrix}
			\displaystyle \frac{1} {\sin(p_n-q_m)}
	\end{bmatrix}} \mathrm{Pf} \begin{bmatrix}
		\widehat K_1(p_m,q_n) & \widehat K_2(p_m,q_n) \\
		-\widehat K_2(p_n,q_m) & \widehat K_3(q_m,q_n)
	\end{bmatrix}_{1 \leq m,n \leq k}, \label{eq:generator}
\end{align}
namely, a ratio of a determinant and a Pfaffian. The latter is built of $2 \times 2$ blocks, consisting of the three kernel functions
\begin{equation}
\begin{split}
	\widehat K_1(p_m,p_n) =& 2N (2N+1) \sin^{2N-1}(p_n-p_m) q_{2N-2}^{(N)} \left (  \frac{\cos(p_n+p_m)}{i \sin(p_n-p_m)}\right ), \\
	\widehat K_2(p_n,q_m) = &\frac{\cos^{2N+1}(p_n-q_m) \cos(p_n+q_m)}{\sin(p_n-q_m) \cos (2p_n)} \\
	&  + (-1)^{N+1}(2N+1)  \frac{\sin^{2N+1}(p_n+q_m) }{\cos(2p_n)} q_{2N}^{(N+1)} \left ( \frac{\cos(p_n-q_m)}{i \sin(p_n+q_m)}\right ), \\
	\widehat K_3 (q_m,q_n) = & (-1)^{N} \sin(q_n-q_m) \cos^{2N+2}(q_n+q_m) \Phi_{2N+2}^{(1)} \left ( \cos^{2} (q_n+q_m) \right )  \\
	& -\sin^{2N+1}(q_n-q_m) q_{2N}^{(N+1)} \left (\frac{\cos (q_m+q_n)}{i\sin(q_m-q_n)} \right ),
\end{split}\label{kernels}
\end{equation}
where we have employed~\cite[Eq.~(21)]{Hahn2023a} in a slightly modified but equivalent form since the expression simplifies drastically with our particular choice of $v(p)$. The special functions involved in~\eqref{kernels} are the Lerch transcendent~\cite{DLMF}
\begin{align}
	\Phi_{n}^{(1)} (x) = -\frac{1}{x^{n}} \left [ \mathrm{log}(1-x) + \sum_{j = 1}^{n-1} \frac{x^{j}}{j}\right]=\sum_{j=0}^\infty\frac{x^j}{j+n} =\frac{1}{n}\hyp{1}{n}{n+1}{x}\label{eq:lerch}
\end{align}
and the skew-orthogonal polynomial $q_{2(N-\alpha)}^{(N+1-\alpha)}( x)$ which is essentially a  hypergeometric function~\cite{DLMF},
\begin{align}
	q_{2(N-\alpha)}^{(N+1-\alpha)}(x) =\sum_{m=0}^{N-\alpha}\frac{B(N+1-\alpha,3/2)}{B(m+1,N-\alpha-m+3/2)}x^{2m}= x^{2(N- \alpha)}\hyp{-N + \alpha}{1}{3/2}{-\frac{1}{x^{2}}} \label{qhyp}
\end{align}
with $\alpha$ a non-negative integer and $B(\nu,\mu)$ the Beta function. We would like to point out that $q_{2(N-\alpha)}^{(N+1-\alpha)}(x)=q_{2(N-\alpha)}^{(N+1-\alpha)}(-x)  $ is even and it is normalised by the identity $q_{2(N-\alpha)}^{(N+1-\alpha)}(\pm i)=(-1)^{N-\alpha}/(1+2N-2\alpha)$.

In Sec.~\ref{sec:kpoint}, we compute the following expression for the $k$-point correlator,
\begin{align}
	&C_{k}(p) = \sum_{\sigma \in \mathbb{S}_{k}} \sum \limits_{m = 0}^{\lfloor k/2 \rfloor}  \frac{1}{(2m)! (k-2m)!} \mathrm{perm} \left [  \frac{1- \delta_{ab}}{\sin(p_{\sigma(a)}-p_{\sigma(b)} ) }\right ]_{1 \leq a,b \leq 2m} \nonumber \\
	& \times\mathrm{Pf} \left [ \begin{bmatrix}
		0 & C_1(p_{\sigma(a)}) \\
		-C_1(p_{\sigma(a)}) & 0
	\end{bmatrix} \delta_{a b}  + ( 1 - \delta_{ab} ) \begin{bmatrix}
		\widehat K_1(p_{\sigma(a)},p_{\sigma(b)}) & \widehat K_2(p_{\sigma(a)},p_{\sigma(b)}) \\
		-\widehat K_2(p_{\sigma(b)},p_{\sigma(a)}) & \widehat K_3(p_{\sigma(a)},p_{\sigma(b)})
	\end{bmatrix} \right]_{2m+1 \leq a,b \leq k}. \label{eq:kpointcorrelator}
\end{align}
The outer summation runs over all elements of the symmetric group $\mathbb{S}_k$ the inner sum goes over products of permanents and Pfaffians with matrix  arguments of the different sizes $2m \times 2m$ and $(k-2m-1) \times (k-2m-1)$. The simplest case is $k=1$, in particular the density
\begin{align}\label{eq:c1punkt}
	C_1(p) =\tan (2p)\left[ 1+(-1)^{N+1}  (2N+1) \sin^{2N} (2p) q_{2N}^{(N+1)} \left ( \frac{1}{i \sin(2p)}\right )\right].
\end{align}
On the first glance it looks that it has singularities at $p=(1+2n)\pi/4$ for any $n\in\mathbb{Z}$. However, the expression in the bracket vanishes at these points, too, because of the normalisation $q_{2N}^{(N+1)}( \pm i )=(-1)^N/(2N+1)$.

For $k=2$, we obtain the two-point correlator
\begin{align}
	C_{2}(p) = &C_1(p_1)C_1(p_2)  - \frac{1 - \cos^{4N+2}(p_1-p_2) }{\sin^{2}(p_1-p_2) } + \frac{1}{\cos (2p_1) \cos (2p_2)}\nonumber \\
	&  \times \left [ \cos^{4N+2}(p_1-p_2) - (2N+1)^{2} \sin^{4N+2}(p_1+p_2) q_{2N}^{(N+1)} \left ( \frac{\cos(p_1-p_2)}{i \sin(p_1+p_2)} \right )^{2}\right ] \nonumber \\
	& -2N (2N+1)  \sin^{4N}(p_1-p_2) q_{2N}^{(N+1)}\left (\frac{\cos(p_1+p_2)}{i\sin(p_1-p_2)} \right ) q_{2N-2}^{(N)}\left (\frac{\cos(p_1+p_2)}{i\sin(p_1-p_2)} \right ) \nonumber \\
	& +  (-1)^{N}2N(2N+1) \cos^{2N+2}(p_1+p_2) \sin^{2N}(p_1-p_2) \Phi_{2N+2}^{(1)} \left ( \cos^{2}(p_1+p_2) \right ) \nonumber \\
	& \times q_{2N-2}^{(N)} \left ( \frac{\cos(p_1+p_2)}{i \sin(p_1-p_2)}\right ),\label{eq:c2punkt}
\end{align} 
where we used the trigonometric identity $\cos^2(p_1+p_2)=\cos(2p_1)\cos(2p_2)+\sin^2(p_1-p_2)$ as well as the symmetry of $q_{2N}^{(N+1)}(x)$. We split the second and third term in this peculiar way as then the singularity at $p_1=p_2$ is regularised. Also the apparent poles at $p_1=(1+2n)\pi/4$ or $p_2=(1+2n)\pi/4$ for any $n\in\mathbb{Z}$ are regularised because of $\cos(\pi/4-x)=\sin(\pi/4+x)$ and the normalisation of the polynomial.

\subsection{First and Second Moment}
\label{sec:secw1}

We compute the first two moments by integrating the corresponding correlator according to Eq.~\eqref{eq:momente}. In case of the first moment it must be
\begin{equation}
	\langle W \rangle = 0
\end{equation}
since the one-point correlation function~\eqref{eq:c1punkt} satisfies the periodicity $C_1(p)=C_1(p+\pi)$ and the antisymmetry $C_1(p)=-C_1(-p)$. Furthermore, it is a continuous function on $[0,2\pi]$ so that no integrability issues arise.

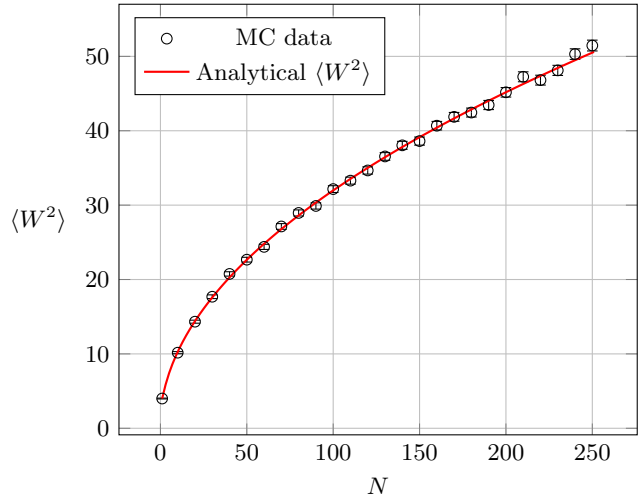
\begin{figure}[h!]
	\centering
	\begin{tikzpicture}
		\begin{axis}[
			xlabel={$N$},
			ylabel={$\langle W^2\rangle$},
			ylabel style={rotate=-90},
			grid=major,
			font=\small,
			legend pos=north west
			]
			\addplot[
			only marks,
			mark=o,
			black,
			error bars/.cd,
			y dir=both,
			y explicit
			] coordinates {\WsqMC};
			\addlegendentry{MC data}
			
			\addplot[
			red,
			thick
			] coordinates {\WsqAna};
			\addlegendentry{Analytical $\langle W^2\rangle$}
		\end{axis}
	\end{tikzpicture}
	\caption{Second moment of winding number for finite $N$: The dimension of the random matrix is on the abscissa and the second moment of the winding number on the ordinate. The points were generated for $N=1, 10, 20,  \dots, 250$, each with $10^4$ realisations of the random matrix model. }
	\label{fig:nummericsWsq}
\end{figure}

For  the second moment $\langle W^{2} \rangle $, we proceed analogously by integrating $C_2$    over the  manifold $[ 0, 2\pi ]^{2} $. As the first moment vanishes it is reasonable to consider only the connected part of the correlation function,
\begin{equation}
	\tilde C_{2}(p_1,p_2) = {C}_{2}(p_1,p_2) - C_{1}(p_1)C_{1}(p_2).
\end{equation}
Thus, it is
\begin{equation}
	\langle W^{2} \rangle = - \frac{1}{4 \pi^{2}} \intl p_1 \intl p_2 \, \tilde{C}_{2}(p_1,p_2). 
\end{equation}
Considering the explicit form of the two-point correlation function~\eqref{eq:c2punkt}, we notice that the center-of-mass and relative coordinates,
\begin{equation}
	\Omega = p_1+p_2 \qquad{\rm and}\qquad \Delta = p_1-p_2, 
\end{equation}
are more suitable. The Jacobian is clearly ${\rm d}p_1{\rm d}p_2={\rm d}\Omega{\rm d}\Delta/2$ while the new domain is $\Omega\in[0,4\pi]$ and $\Delta\in[0,2\pi]$ which can be found by geometric considerations of a two-dimensional torus. Then, we obtain 
\begin{equation}
	\langle W^{2} \rangle = -\frac{1}{8\pi^{2}} \int\limits_0^{4\pi} {\rm d}\Omega \int\limits_0^{2\pi} {\rm d}\Delta  \,\tilde{C}_2\left (\frac{\Omega+ \Delta}{2},\frac{\Omega-\Delta}{2} \right )   
	=  -\frac{1}{8 \pi^{2}} \sum_{i = 1}^{4} D_i .
\end{equation}
The $D_i$ are given as 
\begin{align}
	D_1 = & \int\limits_0^{4\pi} {\rm d}\Omega \int\limits_0^{2\pi}  {\rm d}\Delta \, \frac{\cos^{4N+2} \Delta - 1}{\sin^{2} \Delta} ,\\
	D_2 = & \int\limits_0^{4\pi} {\rm d}\Omega \int\limits_0^{2\pi}  {\rm d}\Delta \frac{\cos^{4N+2}\Delta- (2N+1)^{2} \sin^{4N+2} \Omega \ q_{2N}^{(N+1)}\left( - i \cos \Delta / \sin \Omega\right)^{2}}{\cos(\Omega-\Delta) \cos( \Omega +\Delta) }, \label{eq:defd2} \\
	D_3 = & -2N(2N+1) \int\limits_0^{4\pi} {\rm d}\Omega \int\limits_0^{2\pi}  {\rm d}\Delta \sin^{4N} \Delta \  q_{2N}^{(N+1)} \left ( \frac{\cos \Omega}{i\sin \Delta}\right )q_{2N-2}^{(N)} \left ( \frac{\cos \Omega}{i\sin \Delta}\right ), \label{eq:defd3} \\
	D_4 = &  (-1)^{N}2N (2N+1) \int\limits_0^{4\pi} {\rm d}\Omega \int\limits_0^{2\pi}  {\rm d}\Delta \cos^{2N+2} \Omega\ \sin^{2N} \Delta\ q_{2N-2}^{(N)} \left ( \frac{\cos \Omega}{i\sin \Delta}\right ) \Phi_{2N+2}^{(1)} \left ( \cos^{2} \Omega \right ). \label{eq:defd4}
\end{align}
An integral similar to $D_1$ was already calculated in the context of parametric correlations in the unitary case \cite{Braun2022} which becomes in our case
\begin{equation} 
	D_1 = 8 \pi B(-1/2,2N+3/2).
\end{equation}
It happens that $D_3$ exactly vanishes, see a proof in Appendix~\ref{secd3}, while $D_2$ and $D_4$ are computed in Appendices~\ref{secd2} and~\ref{secd4}.
We recombine the results of $D_i$ in Eq. \eqref{eq:zweitemoment} and obtain the final expression of the second moment 
\begin{align}
	\langle W^{2}\rangle =\frac{2}{\pi} \left[ (-1)^{N+1}B\left(N+\frac{1}{2},N+\frac{1}{2}\right)-B\left(2N+\frac{3}{2},-\frac{1}{2}\right)\right].\label{eq:zweitemoment}
\end{align}
Monte Carlo simulations confirm this result, see Fig.~\ref{fig:nummericsWsq}.

It is now simple to find the asymptotics of $\langle W^{2} \rangle$ in~\eqref{eq:zweitemoment} in the limit $N \to \infty$ where we can use Stirling's formula to obtain
\begin{align}
	B(-1/2,2N+3/2) \overset{N\to\infty}{\sim} -2 \sqrt{2\pi N}\qquad{\rm and}\qquad
	B(N+1/2,N+1/2) \overset{N\to\infty}{\sim} \frac{1}{2^{2N+2}}\sqrt{\frac{\pi}{N}}. \label{eq:secondcontribution}
\end{align}
Hence, the first term in~\eqref{eq:zweitemoment}  is exponentially suppressed so that it is
\begin{equation}
	\langle W^{2} \rangle \overset{N\to\infty}{\sim} \sqrt{\frac{32N}{\pi}}. \label{eq:w2asymptotics}
\end{equation}
This indicates that the winding number as a random variable is scaling with $N^{1/4}$ which agrees with the scaling in the complex case~\cite{Braun2022,Hahn2025}. It is this scale we will use in the subsequent unfolding of the winding number distribution.

\section{The probability distribution of the winding number and its Gaussian limit}
\label{sec:unfolding}

There is also a probabilistic approach via Lyapunov's central limit theorem~\cite{Lyapunov} to show that the limiting probability distribution of the winding number is Gaussian. In Sec.~\ref{sec:dist}, we prepare the ground to apply the theorem. As the coefficient functions $(a(p),b(p))=(\cos(p),i\sin(p))$ in the considered two matrix model $K(p)=\cos(p)K_1+i\sin(p)K_2$ can be traced back to a contour of the unit circle in the complex plane, we can make use of the joint probability density function of the complex eigenvalues for the quaternion spherical ensemble~\eqref{jpdf} which allows us to integrate out the angles of those complex eigenvalues which leave us with a sum of products of probabilities. This follows the same strategy as in~\cite{Braun2022} for the class AIII. In Sec.~\ref{sec:Gauss.limit}, we identify these probabilities to one of a sum of non-identically but independently distributed Bernoulli random variables for which the central limit theorem applies under certain conditions. We show in this sections that those conditions are satisfied.

\subsection{Closed form expression of the winding number probability distribution}
\label{sec:dist}

In the second, AIII-analogous approach, we want to derive a formula for the number probability density function. An exact, compact formula at finite $N$ is, however, elusive, as we will see. We will actually arrive at a sum over a permutation, which is unhandy for large $N$. Yet, it gives us a way to derive the Gaussian distribution for $N\to\infty$, see Sec.~\ref{sec:unfolding}.

An expression for $P(W)$ has been calculated in~\cite{Braun2022} for the chiral unitary case AIII. We aim to do the same for CII. Hence, we briefly sketch the ansatz from~\cite{Braun2022} and apply it to our case. It is significant to recast the random matrix field 
\begin{equation}
	K(p) = \frac{s(p)}{2} (K_1+K_2) +\frac{1}{2s(p)} (K_1 -K_2) \label{eq:distnewmodel},
\end{equation}
by substituting $s=s(p) = e^{ip}$. The sum and difference $K_\pm = K_1\pm K_2$ of the quaternionic Ginibre matrices are also independently Ginibre distributed.
The determinant of $K(p)$ is
\begin{align}
	\det K(p) = \frac{\det K_-}{(2s)^{2N}} \det(s^{2} \mathds{1}_{2N}+K_-^{-1}K_+) = \frac{\det K_-}{(2s)^{2N}} \prod_{n = 1}^{N} (s^{2} +z_{n})(s^{2} + z^{*}_{n}) .\label{eq:distfinaldet2}
\end{align}
with  $z_1,\ldots,z_N$ the eigenvalues of the product matrix $K_-^{-1}K_+$ which are drawn from the quaternionic spherical ensemble~ \cite{Forrester2009,Mays2013,Mays2017}.  Our goal is to connect the model to the defintion of $W$. We utilise Eq.~\eqref{eq:definitionw} and insert the new expression for $\det K(p)$  to find
\begin{equation}
	W = \frac{1}{2 \pi i} \oint \limits_{|s| = 1} \mathrm{d} s \, \frac{d}{ds} \ln \det K(s) = N_{Z} - N_{P}.
\end{equation}
This is Cauchy's argument principle that relates the winding number $W$ to the number of the zeros $N_Z$ and poles $N_P$ counted with multiplicity of $\det K(p)$ inside the unit circle $|s| = 1$, which effectively makes $W$ solely dependent on the distribution of the eigenvalues $z_n$. It is evident from Eq.~\eqref{eq:distfinaldet2} that the number of poles counted with multiplicity is given simply by the one at the origin which is $N_P=2N$. The zeros come in complex conjugate pairs so that the number $m= 0,\dots,N$ of zeros (without counting multiplicity) above the real line and inside the unit circle gives a winding number (apart from a set of measure zero) equal to
\begin{equation}
	W = 4m -2N \quad \text{with} \quad W = -2N,\dots,0,\dots 2N. \label{eq:wnullstellenpole}
\end{equation}
In Appendix~\ref{sec:rm}, we explicitly calculate the probability $r(m)$ to find a specific number $m$ of eigenvalue pairs inside the unit discs $D(0,1)$. We do this by integrating the joint eigenvalue probablility density function 
\begin{equation}\label{jpdf}
	G^{(4)}(\omega) = \frac{1}{c^{(4)}} \Delta_{2N}(\omega)\prod_{n = 1}^{N} \frac{z_n - z_n^{*}}{(1+|z_{n}|^{2})^{2N+2}}, \quad c^{(4)} = (2\pi)^{N} N! \prod \limits_{n = 1}^{N} B(2n,2N+2-2n)
\end{equation}
with $\omega = \mathrm{diag}(z_1,z_1^{*},\dots,z_N, z_N^{*} )$ from \cite{Mays2017} over the region $D^m(0,1)\times [\mathbb{C}\setminus D(0,1)]^{N-m}$,
\begin{equation}
	r(m) = \int \limits_{|z_1| \leq 1} \mathrm{d} z_1 \dots  \int \limits_{|z_m| \leq 1} \mathrm{d} z_{m} \int \limits_{|z_{m+1} | > 1} \mathrm{d} z_{m+1} \dots \int \limits_{|z_{N}| > 1} \mathrm{d} z_{N}\, G^{(4)}(\omega).\label{eq: r integrals}
\end{equation}
We employ the Vandermonde determinant
\begin{equation}
	\Delta_{2N}(\omega) = \det \begin{bmatrix} \omega_{m}^{n-1}\end{bmatrix}_{1 \leq n,m \leq 2N}.
\end{equation}
The probability $r(m)$ is then the permanent
\begin{equation}
	r(m) = \frac{1}{N!}\sum_{\sigma \in \mathbb{S}_{N}} \left (\prod_{n = 1}^{m} p_{\sigma(n)} \right ) \left (\prod_{n = m+ 1}^{N} [1-p_{\sigma(n)}] \right )\label{eq: r(m)}
\end{equation}
with the incomplete normalised beta functions, following the ideas of~\cite{Braun2022},
\begin{align}\label{incomplete.Beta}
	p_m= \frac{2}{B(2m,2N-2m+2)} \int \limits_{0}^{1}  \frac{x^{4m-1}\mathrm{d} x }{(1+x^{2})^{2N+2}} \ {\rm and}\  1-p_m=& \frac{2}{B(2m,2N-2m+2)} \int \limits_{1}^{\infty}  \frac{x^{4m-1}\mathrm{d} x}{(1+x^{2})^{2N+2}}.
\end{align}
We notice that permutation invariance of the eigenvalues has not been incorporated, yet, cf.~\eqref{eq: r integrals}. Thus, once we symmetrise in $z_1,\ldots,z_N$ one obtains a binomial prefactor and we arrive at the winding number probability
\begin{equation}
	P(W) = \binom{N}{(W+2N)/4} r \left ( \frac{W+2N}{4}\right ),
\end{equation}
which we will evaluate in the large $N$ limit in Sec.~\ref{sec:Gauss.limit}.

\subsection{Gaussian limit of the winding number probability}\label{sec:Gauss.limit}

 In contrast to the correlator, discussed in the next subsection, we use the insight from~\eqref{eq:w2asymptotics}, which is similar to the results in~\cite{Braun2022} for the class AIII, and unfold $W$ according to
\begin{equation}
	W = \Xi N^{1/4}=4m -2N
\end{equation}
with a rescaled winding number $\Xi$. Hence, we consider $P(W)$ on this new scale
\begin{equation}
	P(\Xi) = \binom{N}{(\Xi N^{1/4} + 2N)/4} r \left ( \frac{\Xi N^{1/4} + 2N}{4} \right ) , \label{eq:newdist}
\end{equation}
where we slightly abuse notation by employing the same symbol for the probability function of $W$ and $\Xi$.

We use a probabilistic approach to calculate the limit of 
\begin{align}
	\lim \limits_{N \to \infty} \frac{P(\Xi)}{P(0)} = \lim \limits_{N \to \infty} \frac{\displaystyle \binom{N}{(\Xi N^{1/4} + 2N)/4} r \left ( \frac{\Xi N^{1/4} + 2N}{4} \right ) }{\displaystyle \binom{N}{N/2}r \left ( \frac{N}{2}\right )} .\label{eq:normalisedratio}
\end{align}
First, we recall that $r(m)$ describes the probability~\eqref{eq: r(m)} of finding $m$ (out of a total of $N$) distinguishable eigenvalues of the matrix $Y=K_-^{-1}K_+$ (drawn from the quaternionic spherical ensemble) inside the unit circle, where the eigenvalues appear in complex conjugate pairs due to the quaternion structure. We recall that the binomial coefficient appears when passing to indistinguishable eigenvalues. Consequently, we sum over subsets of $\{1,\dots,N\}$ with cardinality $m$. For sake of clearness, we keep the abbreviation $m=(\Xi N^{1/4} + 2N)/4$ in our equations. When expanding the permanent in~\eqref{eq: r(m)} we arrive at the result 
\begin{align}
	P(\Xi)&=\binom{N}{m} r(m) =\sum_{\substack{S\subset\{1,\dots,N\}\\|S|=m}} \,\prod_{k\in S}p_k\, \prod_{k\notin S}(1-p_k),
\end{align}
see~\eqref{incomplete.Beta}.

Each eigenvalue can either lie inside or outside the unit circle, thus motivating the use of Bernoulli distributions. The $k$'th eigenvalue is described by the random variable $X_k$ which is one, if the eigenvalue lies inside, and zero if it lies outside the unit circle. These $X_k$ follow Bernoulli distributions. The probabilities of the $k$'th eigenvalue lying inside (event $X_k=1$)  and outside (event $X_k=0$) the unit circle is then
\begin{align}
	\mathbb{P}(X_k=1)=p_k,\quad\mathbb{P}(X_k=0)=1-p_k.
\end{align}
It is a map to a binary event where each random variable has its own independent probability. Due to its binary structure it is for any moment $\langle X_k^j\rangle=p_k$ for all $j>0$ and $k=1,\ldots,N$.

It is important to notice, that unlike the fixed probabilities in the standard central limit theorem, these probabilities are $N$ dependent. Hence, the sum over the subsets $S\subset\{1,\dots,N\}$ in the definition of $P(\Xi)$ can be interpreted as a probability
\begin{gather} 
	P(\Xi)=\sum_{\substack{S\subset\{1,\dots,N\}\\|S|=m}}\prod_{k\in S}p_{k} \prod_{k\notin S}(1-p_k)=\sum_{|S|=m}\mathbb{P}(S)=\mathbb{P}\left(\sum_{k=1}^N X_k=m\right).
\end{gather}
This probabilistic problem calls for a central limit theorem when taking the limit of large $N$. However, the standard central limit theorem does not apply as the random variables $X_l$ are independent yet not identically distributed.

We employ the more general Lyapunov central limit theorem~\cite{Lyapunov}, which allows for probabilities that are $N$ dependent and applies to sums of independent random variables. It states that the sum of random variables recentred by the sum of means $\mu_N$ and rescaled by the sum of variances $s_N$ converges in distribution to a normal distribution, if the rescaled and recentered $2+\varepsilon$'th moment vanishes for some $\varepsilon>0$ in the limit $N\to\infty$, i.e.,
\begin{align}\label{cond.Lyap}
	&\frac{1}{s_N^{2+\varepsilon}}\left\langle \sum_{k=1}^N(X_k-\langle X_k\rangle)^{2+\varepsilon}\right\rangle \leq \frac{1}{s_N}\overset{N\to\infty}{\longrightarrow} 0 \implies\frac{\sum_{k=1}^N X_k-\mu_N}{s_N}\overset{d}{\to} \mathcal{N}(0,1).
\end{align}
In our case it is
\begin{align}
	&\mu_N = \sum_{k=1}^N\langle X_k\rangle= \sum_{k=1}^N p_k\quad{\rm and}\quad s_N^2 = \sum_{k=1}^N\sigma_k^2= \sum_{k=1}^N\langle (X_k-p_k)^2\rangle = \sum_{k=1}^N p_k(1-p_k),
\end{align}
which satisfies the Lyapunov condition as we will see. 

The mean value can be computed directly from~\eqref{incomplete.Beta},
\begin{equation}
\begin{split}
\mu_N=&\sum_{k=1}^N\frac{2(2N+1)!}{(2k-1)!(2N-2k+1)!}\int\limits_0^1\frac{x^{4k-1}{\rm d}x}{(1+x^2)^{2N+2}}=\frac{2N+1}{2}\sum_{j=0}^{2N}\binom{2N}{j}\int\limits_0^1\frac{[1-(-1)^j]y^{j}{\rm d}y}{(1+y)^{2N+2}}\\
=&\frac{2N+1}{2}\int\limits_0^1\frac{[(1+y)^{2N}-(1-y)^{2N}]{\rm d}y}{(1+y)^{2N+2}}=\frac{N}{2}.
\end{split}
\end{equation}
In the first line, we substituted $y=x^2$ and generate the sum over $k$ via the auxiliary sum over $j$ by cancelling all even $j$. In the second line we carried out the binomial sums and finally one can integrate directly.

For the sum of variances we need to compute the asymptotic of
\begin{equation}
s_N^2=4(2N+1)^2\sum_{k=1}^N\binom{2N}{2k-1}^2\int \limits_0^1{\rm d}x_1\int \limits_1^\infty {\rm d}x_2\frac{(x_1x_2)^{4k-1}}{(1+x_1^2)^{2N+2}(1+x_2^2)^{2N+2}}.
\end{equation}
We anew substitute $y_l=x_l^2$ for both $l=1,2$ and extend the sum to a summing index $j=0,\ldots,2N$ where the even summands are cancelled by and additional factor $1-(-1)^j$. One of the binomials can be rewritten as the contour integral
\begin{equation}
\binom{2N}{j}=\oint \limits_{|z|=1}\frac{(1+z)^{2N}{\rm d}z}{2\pi i z^{j+1}}
\end{equation}
so that the sum can be carried out,
\begin{equation}\label{sn.eq}
s_N^2=\frac{(2N+1)^2}{2}\int \limits_0^1{\rm d}y_1\int \limits_1^\infty {\rm d}y_2\oint \limits_{|z|=1}\frac{{\rm d}z}{2\pi i z}\frac{(1+y_2z)^{2N}[(1+y_1/z)^{2N}-(1-y_1/z)^{2N}]}{(1+y_1)^{2N+2}(1+y_2)^{2N+2}},
\end{equation}
where we additionally rescaled $z\to y_2z$
We perform a saddle point analysis for each of the two terms.

The first term in~\eqref{sn.eq} with $(1+y_1/z)^{2N}$ has a maximum at $z=\sqrt{y_1/y_2}$ which is only possible at the boundary point $y_1=y_2=1$ and $z=1$. Accounting for this joint saddle point and especially that $z$ crosses the real axis orthogonally, as well as $y_1\leq1$ and $y_2\geq1$, we employ the expansion
\begin{equation}
y_1=1-\frac{\delta y_1}{\sqrt{N}},\quad y_2=1+\frac{\delta y_2}{\sqrt{N}},\quad z=1+\frac{i\delta z-(y_1+y_2)/2}{\sqrt{N}}
\end{equation}
with $\delta y_1,\delta y_2>0$ and $\delta z\in\mathbb{R}$ to find the limit
\begin{equation}
\begin{split}
&\frac{(2N+1)^2}{2}\int \limits_0^1{\rm d}y_1\int \limits_1^\infty {\rm d}y_2\oint \limits_{|z|=1}\frac{{\rm d}z}{2\pi i z}\frac{(1+y_2z)^{2N}(1+y_1/z)^{2N}}{(1+y_1)^{2N+2}(1+y_2)^{2N+2}}\\
\overset{N\to\infty}{\sim}&2N^2\int \limits_0^\infty\frac{{\rm d}\delta y_1}{\sqrt{N}}\int \limits_0^\infty \frac{{\rm d}\delta y_2}{\sqrt{N}}\int \limits_{-\infty}^\infty\frac{{\rm d}\delta z}{2\pi\sqrt{N}}\frac{1}{16}\exp\left[-\frac{\delta z^2}{2}-\frac{(\delta y_1+\delta y_2)^2}{8}\right]=\sqrt{\frac{N}{8\pi}}.
\end{split}
\end{equation}

The second term in~\eqref{sn.eq} with $(1-y_1z)^{2N}$ has no critical point at all. Hence the maximum must be at the boundaries. For a fixed $y_1$ and $y_2$ the maxima are given by $z=\pm i\sqrt{y_1/y_2}$ as can be readily checked for the relevant terms $(1+y_2z)^{2N}(1-y_1/z)^{2N}$. The modulus of the exponential part of the integrand is then
\begin{equation}
\left|\frac{(1+y_2z)^{2N}(1-y_1/z)^{2N}}{(1+y_1)^{2N+2}(1+y_2)^{2N}}\right|_{z=\pm i\sqrt{y_1/y_2}}=\left(\frac{1+y_1y_2}{(1+y_1)(1+y_2)}\right)^{2N}.
\end{equation}
The maximum of this term for $(y_1,y_2)\in[0,1]\times[1,\infty)$ is the boundary given by $(1-y_1)(y_2-1)=0$, meaning one of the two variables $y_1$ and $y_2$ is equal to $1$. The value of the maximum is $2^{-2N}$ while the remaining function is still absolutely integrable for the whole boundary due to the factor $1/[(1+y_1)(1+y_2)]^2$. This means it is
\begin{equation}
\begin{split}
&\left|\frac{(2N+1)^2}{2}\int \limits_0^1{\rm d}y_1\int \limits_1^\infty {\rm d}y_2\oint \limits_{|z|=1}\frac{{\rm d}z}{2\pi i z}\frac{(1+y_2z)^{2N}(1+y_1/z)^{2N}}{(1+y_1)^{2N+2}(1+y_2)^{2N+2}}\right|\\
\leq&\frac{(2N+1)^2}{2}\int \limits_0^1{\rm d}y_1\int \limits_1^\infty {\rm d}y_2\oint \limits_{|z|=1}\frac{|{\rm d}z|}{2\pi}\frac{2^{-2N}}{(1+y_1)^{2}(1+y_2)^{2}}\overset{N\to\infty}{\longrightarrow}0.
\end{split}
\end{equation}
Thus, we have in total for the variance~\eqref{sn.eq}
\begin{equation}
s_N^2\overset{N\to\infty}{\sim}\sqrt{\frac{N}{8\pi}}.
\end{equation}
The condition~\eqref{cond.Lyap} for the Lyapunov central limit theorem is trivially satisfied for $\varepsilon=1$ as $|X_k-\langle X_k\rangle|^3\leq 2|X_k-\langle X_k\rangle|^2$ as $X_k\in\{0,1\}$ are Bernoulli random variables.

\begin{figure}[h]
	\centering
	\begin{tikzpicture}
		\begin{axis}[
			xlabel={$\Xi$},
			ylabel={$\displaystyle \frac{P(\Xi)}{P(0)}$},
			ylabel style={rotate=-90},
			font=\small,
			grid=major,
			legend pos=north west
			]
			\addplot[
			only marks,
			mark=o,
			black,
			error bars/.cd,
			y dir=both,
			y explicit
			] table [x=x, y=y, y error expr={sqrt(\thisrow{y}/9312)}] {
				x y
				0.84 0.9 
				0. 1. 
				2.52 0.38 
				-0.84 0.9 
				-2.52 0.38 
				1.68 0.64 
				-1.68 0.65 
				-4.2 0.06 
				4.2 0.07 
				3.36 0.18 
				-3.36 0.17 
				5.89 0.01 
				5.05 0.02 
				-5.05 0.02 
				-5.89 0. 
				-7.57 0. 
				6.73 0. 
				-8.41 0. 
				7.57 0.
			};
			\addlegendentry{MC data}
			\addplot[
			red,
			thick,
			domain=-10:10,
			samples=100
			] { exp(- x^2 * sqrt(2 * pi) / 16 )};
			\addlegendentry{Analytical $P(\Xi)$}
			\node[draw, fill=white, align=left, anchor=north east, inner sep=4.5pt] at (rel axis cs: 0.97, 0.97) {
				$\mu = 0.02$ \\
				$\sigma^2 = 3.22$ \\
				$\gamma_1 = 0.06$ \\
				$\kappa = 2.94$
			};
			\legend{}
		\end{axis}
	\end{tikzpicture}
	\caption{Unfolded winding number distribution: On the abscissa is the unfolded winding number $\Xi$ and on the ordinate the normalised distribution thereof. The red line is the analytical result for the unfolded distribution whereas the data points were generated in a Monte Carlo simulation ($N=512$ and $5\times10^4$ realisations of the random matrix model). The inset box contains the values for the mean $\mu$, the variance $\sigma^2$, the skewness $\gamma_1$ and the kurtosis $\kappa$. }
	\label{fig:nummericsxi}
\end{figure}
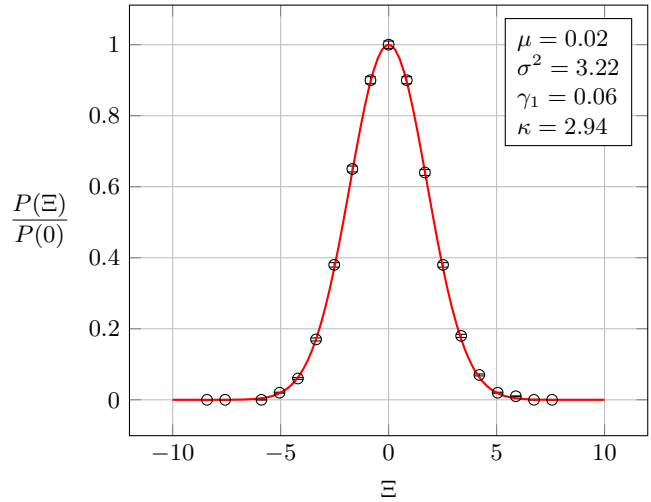

In summary, the probability distribution of the random variables $\sum_{k=1}^NX_k$ becomes Gaussian,
\begin{align}
	&\mathbb{P}\left(\sum_{k=1}^N X_k=m\right)=
	\frac{1}{\sqrt{2\pi s_N^2}}\exp\left(-\frac{(m-\mu_N)^2}{2s_N^2}\right)\\
	&\frac{P(\Xi)}{P(0)}\overset{\text{CLT}}{=} \exp\left[-\sqrt{2\pi}\frac{\Xi^2}{16}\right]
\end{align}
where we recall the relation $W=\Xi N^{1/4}=4m -2N$. The result is in good agreement with numerical simulations as illustrated in Fig.~\ref{fig:nummericsxi}.

The argument of the exponential function contains a $\sqrt \pi/\beta^2$, as in the AIII case \cite{Braun2022}. The new factor $\sqrt{2}$, which does not exist in the AIII case, is explained by twice the number of eigenvalues for a quaternion matrix.

\section{Unfolding of the correlators and the large $N$ limit}\label{sec:unfold}

Random Matrix Theory provides not only universal results for integrated quantities but also for spectral correlations on the scale of the local mean level spacing. Thus, empirical data as well as analytical results have to be unfolded.  Recent works \cite{Braun2022,Hahn2023a,Hahn2023b,Hahn2025} discussed universal features in the chiral unitary case AIII by zooming into parametric correlations.  We want to address the question whether one can find universality in the chiral symplectic case CII, too.

The parameters (e.g., wave vectors) are unfolded according to
\begin{equation}
	p_j = \bar{p}+\frac{\psi_j}{\sqrt{N}}.
\end{equation}
where $ \bar{p}\in[0,2\pi[$ is the base point where one zooms in. The base point can be seen as centre of mass so that the relative coordinates $\psi_j$ sum up to zero which can be enforced by a Dirac delta function.
Therefore, the unfolded $k$-point correlators are defined as the leading large $N$ term of
\begin{equation}\label{fk.def}
\begin{split}
	C_{k} \left ( \bar{p}+ \frac{\psi_1}{\sqrt{N}} , \ldots , \bar{p}+\frac{\psi_k}{\sqrt{N}}\right ) \delta\left(\sum_{j=1}^k\frac{\psi_j}{\sqrt{N}}\right)\left(\prod_{j=1}^k\frac{\mathrm{d} \psi_j}{\sqrt{N}}\right) \mathrm{d}\bar{p}.
\end{split}
\end{equation}
For $k=1$ there is no relative coordinate $\psi$ though at $\bar{p}=l\pi/2$ with $l\in\mathbb{Z}$ one may replace $\bar{p}\to l\pi/2+\psi/\sqrt{N}$.

We note that there is a conjugation symmetry, in particular that eigenvalues must appear in complex (anti-)conjugate pairs, when the matrix $K(p)=\cos p\, K_1+i\sin p\, K_2$ is quaternion. This happens at the points $\bar{p}=0,\pi/2,\pi,3\pi/2$. Yet when $\bar{p}\in[0,2\pi[\setminus\{0,\pi/2,\pi,3\pi/2\}$ the generic eigenvalues of $K(\bar{p})$ are not related, though there is evidently a relation to those of $K(-\bar{p})=(\sigma_2\otimes \mathds{1})K^*(\bar{p})(\sigma_2\otimes \mathds{1})$ and $K(\bar{p})=-K(\bar{p}+\pi)$; the latter is due to the specific model. Thence, we should find different behaviours at these different classes.

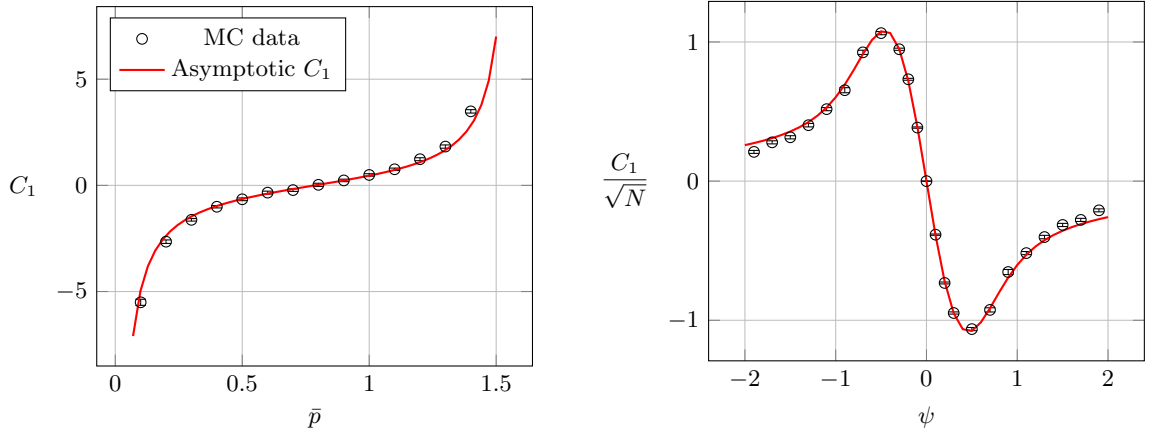
\begin{figure}[htbp]
	\centering
	\begin{subfigure}[b]{.48\textwidth}
		\centering
		\begin{tikzpicture}
			\begin{axis}[
				width=\linewidth,
				trig format plots=rad,
				xlabel={$\bar{p}$},
				ylabel={$C_1$},
				ylabel style={rotate=-90},
				grid=major,
				font=\small,
				legend pos=north west
				]
				\addplot[
				only marks,
				mark=o,
				black,
				error bars/.cd,
				y dir=both,
				y explicit
				] coordinates {\COneBulkMC};
				\addlegendentry{MC data}
				
				\addplot[
				red,
				thick,
				domain=0.07:1.5,
				samples=50
				] {- cot(2 * x)};
				\addlegendentry{Asymptotic $C_1$}
			\end{axis}
		\end{tikzpicture}
	\end{subfigure}
	\hfill
	\begin{subfigure}[b]{.48\textwidth}
		\centering
		\begin{tikzpicture}
			\begin{axis}[
				width=\linewidth,
				xlabel={$\psi$},
				ylabel={$\displaystyle\frac{C_1}{\sqrt{N}}$},
				ylabel style={rotate=-90},
				grid=major,
				font=\small,
				legend pos=north west
				]
				\addplot[
				only marks,
				mark=o,
				black,
				error bars/.cd,
				y dir=both,
				y explicit
				] coordinates {\COneSpecMC};
				\addlegendentry{MC data}
				
				\addplot[
				red,
				thick
				] coordinates {\COneSpecAna};
				\addlegendentry{Asymptotic $\frac{C_1}{\sqrt{N}}$}
				\legend{}
			\end{axis}
		\end{tikzpicture}
	\end{subfigure}
	\caption{One-point correlator in large $N$ limit: The data points were obtained from a Monte Carlo simulation with $10^5$ realisations and $N=50$. \textbf{Left}: One-point correlator for bulk base points $\bar{p}$ (on the abscissa). \textbf{Right}: Unfolded one-point correlator for base point $\bar{p}=\pi/2$, plotted over the relative coordinate $\psi$. The solid red curves are the formulae from Eq.~\eqref{eq:f1result}. }
	\label{fig:nummericsC1}
\end{figure}

Taking these considerations into account, we find the limit for the $1$-point correlator~\eqref{eq:c1punkt},
\begin{equation}
C_1 \left (\bar{p}+\frac{\psi}{\sqrt{N}}\right)\overset{N\to\infty}{\sim}\left\{\begin{array}{cl} \displaystyle-\sqrt{\pi N} \mathrm{erfi}(2\psi)e^{-4\psi^2}, & \displaystyle \bar{p}=\frac{l\pi}{2}\ {\rm with}\ l\in\mathbb{Z},\\ \displaystyle -\cot(2\bar{p}), & \displaystyle \bar{p}\in\mathbb{R}\setminus\frac{\pi}{2}\mathbb{Z}\end{array}\right.\label{eq:f1result}
\end{equation}
following from the limits~\eqref{q.lim.a} and~\eqref{q.lim.b} of the skew-orthogonal polynomial $q_{2N}^{N+1}(-1/\sin(2p))$. The function ${\rm erfi}(y)=2/\sqrt{\pi}\int \limits_0^x\exp[y^2]{\rm d}y$ is the imaginary error function. The result is in good agreement with simulation data, cf. Fig.~\ref{fig:nummericsC1}

The same can be done for the connected two-point correlator~\eqref{eq:c2punkt}. For this quantity we need to distinguish actually $\bar{p}\in\mathbb{R}\setminus\pi/4\mathbb{Z}$, $\bar{p}\in\pi/2\mathbb{Z}$ and $\bar{p}\in\pi/2\mathbb{Z}+\frac{\pi}{4}$, when taking the limit $N\to\infty$. At the zeros of the one-point correlator at $\bar{p} =\pi(2l+1)/4$ with $l\in\mathbb{Z}$ the way how the limit is approached is slightly different though the limit is the same as for $\bar{p}\in\mathbb{R}\setminus\pi/4\mathbb{Z}$. This is mathematically visible in the various asymptotic of the skew-orthogonal polynomials $q_{2N-2a}^{N+1-a}$, see~\eqref{q.lim.a} and~\eqref{q.lim.b} and the argument $\cos(p_1+p_2)/\sin(p_1+p_2)$ in~\eqref{eq:c2punkt} which can diverge as well as converge to zero for these base points. An explanation of such points might be that at $\bar{p}\in\pi/2\mathbb{Z}+\pi/4$ the spectrum of the matrix $K(\bar{p})=(\pm K_1\pm i K_2)/\sqrt{2}$ is perfectly isotropic as it is equal to the complex Ginibre ensemble. Thus these points have higher symmetry. Yet, in the end it has not an impact on the limiting behaviour.

For $\bar{p}\in\mathbb{R}\setminus\pi/4\mathbb{Z}$, meaning for the bulk of parameters, it is
\begin{equation}
\left(\frac{\cos(p_1-p_2)}{i \sin(p_1+p_2)}\right)^2\overset{N\to\infty}{\sim}-\frac{1}{\sin^2(\bar{p})}<-\frac{1}{2},\qquad \left(\frac{\cos(p_1+p_2)}{i \sin(p_1-p_2)}\right)^2\overset{N\to\infty}{\sim}-\frac{N\cos^2(2\bar{p})}{(\psi_1-\psi_2)^2},
\end{equation}
and $\cos^2(p_1+p_2)\overset{N\to\infty}{\sim}\cos^2(2\bar{p})<1$. Applying the respective limits in~\eqref{q.lim.a},~\eqref{q.lim.b} and~\eqref{phi.lim}, we arrive at the asymptotic
\begin{equation}\label{cor2.lim.a}
\begin{split}
\frac{C_1(p_1)C_1(p_2)-C_{2}(p)}{N} \overset{N\to\infty}{\sim} &   \frac{1 - \exp[-2(\psi_1-\psi_2)^2] }{(\psi_1-\psi_2)^2 }=:C_2^{\rm bulk}(\psi_1,\psi_2) .
\end{split}
\end{equation}
The second term in~\eqref{eq:c2punkt} is algebraically suppressed by a factor $1/N$ while the third and fourth term are exponentially small due to factors of the form $\cos^{4N}(2\bar{p})$. This is the same result (up to a factor $2$ due to twice the number of eigenvalues) which was also found in~\cite{Braun2022,Hahn2025} for complex case (class AIII). Therefore, in the bulk of parameters we have some kind of super-universality which does not distinguish between symmetry classes.

We find the same result~\eqref{cor2.lim.a} also for $\bar{p} =\pi(2l+1)/4$ with $l\in\mathbb{Z}$ however the rates of convergences of the other three last terms in~\eqref{eq:c2punkt} are different. While the second term in~\eqref{eq:c2punkt} is still vanishing like $1/N$ it is because the two terms in the parentheses cancel in leading order when plugging in $\cos^2(p_1-p_2),\sin^2(p_1+p_2)\sim1$ for $N\to\infty$. The third and fourth term in~\eqref{eq:c2punkt} vanish like $N^{1-2N}e^{\gamma N}$ with some real constant $\gamma$ after plugging in the asymptotic~\eqref{q.lim.a},~\eqref{q.lim.b} and~\eqref{phi.lim}.

When $\bar{p}\in\pi/2\mathbb{Z}$, we find a different limit. For these points, we use
\begin{equation}
\left(\frac{\cos(p_1-p_2)}{i \sin(p_1+p_2)}\right)^2\overset{N\to\infty}{\sim}-\frac{N}{(\psi_1+\psi_2)^2},\qquad \left(\frac{\cos(p_1+p_2)}{i \sin(p_1-p_2)}\right)^2\overset{N\to\infty}{\sim}-\frac{N}{(\psi_1-\psi_2)^2},
\end{equation}
as well as
\begin{equation}
\cos^2(p_1\pm p_2)\overset{N\to\infty}{\sim}\left(1-\frac{(\psi_1\pm\psi_2)^2}{2N}\right)^2,\qquad \sin^2(p_1\pm p_2)\overset{N\to\infty}{\sim}\frac{(\psi_1\pm\psi_2)^2}{N}.
\end{equation}
Employing~\eqref{q.lim.a},~\eqref{q.lim.b} and~\eqref{phi.lim}, we arrive at
\begin{equation}
\begin{split}
\frac{C_1(p_1)C_1(p_2)-C_{2}(p)}{N}\overset{N\to\infty}{\sim}  \frac{1 - \exp[-2(\psi_1-\psi_2)^2] }{(\psi_1-\psi_2)^2 }+\pi [{\rm erfi}^2(\psi_1+\psi_2)-{\rm erfi}^2(\psi_1-\psi_2)]e^{-4(\psi_1^2+\psi_2^2)}& \\
	\quad +\sqrt{4\pi}(\psi_1-\psi_2){\rm erfi}(\psi_1-\psi_2)\Gamma[0,(\psi_1+\psi_2)^2]e^{-2(\psi_1^2+\psi_2^2)}=:C_2^{\rm crit}(\psi_1,\psi_2)&,
\end{split}\label{eq:f2result}
\end{equation}
where  $\Gamma[0,\widetilde x]=\int \limits_{\widetilde x}^\infty e^{-t}{\rm d}t/t$ is the incomplete gamma function \cite{DLMF}. 

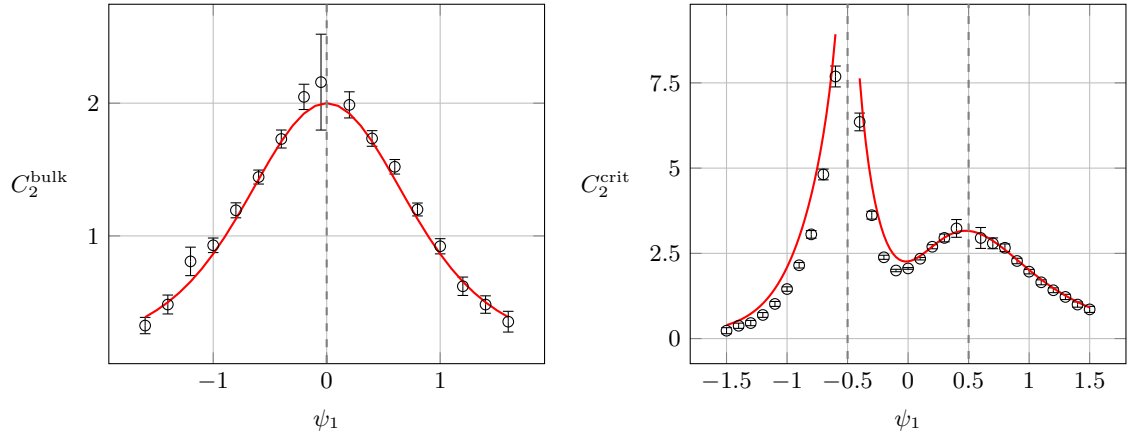
\begin{figure}[htbp]
	\centering
	\begin{subfigure}[b]{.48\textwidth}
		\centering
		\begin{tikzpicture}
			\begin{axis}[
				width=\linewidth,
				trig format plots=rad,
				xlabel={$\psi_1$},
				ylabel={$C_2^{\rm bulk}$},
				ylabel style={rotate=-90},
				grid=major,
				font=\small,
				legend pos=north west
				]
				\addplot[
				only marks,
				mark=o,
				black,
				error bars/.cd,
				y dir=both,
				y explicit
				] coordinates {\CTwoBulkMC};
				\addlegendentry{MC data}
				
				\addplot[
				red,
				thick
				] coordinates {\CTwoBulkAna};
				\addlegendentry{Asymptotic $\frac{C_1}{\sqrt{N}}$}
				\draw [thick, dashed, gray] (axis cs:0, -2) -- (axis cs:0, 16);
				\legend{}
			\end{axis}
		\end{tikzpicture}
	\end{subfigure}
	\hspace{1pt}
	\begin{subfigure}[b]{.48\textwidth}
		\centering
		\begin{tikzpicture}
			\begin{axis}[
				width=\linewidth,
				xlabel={$\psi_1$},
				ylabel={$C_2^{\rm crit}$},
				ylabel style={rotate=-90},
				grid=major,
				font=\small,
				xtick={-1.5, -1, -0.5, 0, 0.5, 1, 1.5},
				ytick={0, 2.5, 5, 7.5},
				legend pos=north west
				]
				\addplot[
				only marks,
				mark=o,
				black,
				error bars/.cd,
				y dir=both,
				y explicit
				] coordinates {\CTwoSpecMC};
				\addlegendentry{MC data}
				
				\addplot[
				red,
				thick,
				unbounded coords=jump
				] coordinates {\CTwoSpecAna};
				\addlegendentry{Asymptotic $\frac{\tilde{C}_2}{N}$}
				
				\draw [thick, dashed, gray] (axis cs:-0.5, -2) -- (axis cs:-0.5, 16);
				\draw [thick, dashed, gray] (axis cs:0.5, -2) -- (axis cs:0.5, 16);
				\legend{}
			\end{axis}
		\end{tikzpicture}
	\end{subfigure}
	\caption{Connected two-point correlator in large $N$ limit: We used the same Monte Carlo dataset ($10^5$ realisations and $N=50$) as in Fig.~\ref{fig:nummericsC1}. The analytical results from Eqs.~(\ref{eq:f1result}, \ref{eq:f2result}) are shown as solid red curves. \textbf{Left}: Bulk base point $\bar{p}=3\pi/8$, and relative coordinate $\psi_2\equiv 0$.  The point $\psi_1=0$ was slightly shifted to $-0.05$ to avoid distortion caused by error bars. \textbf{Right}: Base point $\bar{p}=\pi/2$, and relative coordinate $\psi_2\equiv 0.5$. The points at $\psi_1=\pm 0.5$ are omitted and the analytical curve is truncated near the lograrithmic singularity at $\psi_1=-0.5$ on the interval $[-0.6,-0.4]$. The simulated points lying under the curve around the singularity is a finite $N$ effect, since the data does not diverge at the singularity.}
	\label{fig:nummericsC2}
\end{figure}

We compare the results to simulation data in Fig.~\ref{fig:nummericsC2}.
At the diagonal point $p_1=p_2$, the simulation calculates the autocorrelation, which has a significantly increased variance across the $10^5$ realisations. To avoid this spike in variance, which would distort the $y$-scaling, the point $\psi_1=0$ in the left panel was slightly shifted to $-0.05$. For the same reason, the point $\psi_1=\psi_2=0.5$ was omitted in the right panel. We chose $\psi_2\equiv 0.5$ to compare the simulation to the full profile of the two-point correlator that diverges at $\psi_1=-\psi_2$, due to the logarithmic singularity of the incomplete gamma function $\Gamma[0,x]$ as $x\to0$. This divergence necessitates a different scaling around $\psi_1+\psi_2=0$. The corresponding Monte Carlo point was omitted as well, because it is an order of magnitude larger than the other points and carries substantial uncertainty which would distort the scaling of the $y$-axis. Since the simulated matrices cannot generate a divergence for finite matrix dimension $N$, the points around the singularity lie slightly below the curve.

\newpage
\section{Conclusion}\label{sec:conclusion}

We investigated the statistical aspects of the winding number, which emerges as a topological index in the presence of chiral symmetry. In condensed matter physics the winding number has real physical interpretation and is the topological index for different topological phases. We equipped a chiral symplectic Hamiltonian $H(p)$ drawn from the class CII with a parametric dependence $p$ which can be seen as a wave vector dependence. To keep it simple we considered a matrix model similar to the previous work~\cite{Braun2022}.

Our focus was the discrete probability distribution of the winding number but also the asymptotic of the one- and two-point correlators of the winding number density, from which we could compute compact expressions for the first two moments at finite matrix dimension. For this purpose, we computed the first two correlators explicitly and found that the asymptotic result is up to a scaling the same as for the class AIII as long as the parameter $p$ is such that the off-diagonal block $K(p)$ of $H(p)$ is not a small perturbation of a quaternion matrix. This is true for most $p$, and this result can be anticipated. Only when $p$ is chosen such that $K(p)$ is almost a quaternionic matrix, the correlators change and show another kind of statistics. It has to be checked with other models whether this profile is also universal.

In previous works~\cite{Braun2022,Hahn2023a,Hahn2023b,Hahn2025,Yahiaoui2025}, we addressed the question of universality in the context of the correlators and the winding number distribution in the class AIII, which have been found to exhibit universality on appropriate scales, when taking the limit of infinite matrix dimension. Results in the class AIII have shown the same scaling for the winding number distribution $W = \Xi N^{1/4}$ as it is also observed here in the class CII.

It is the second approach which sheds light on where the rather unusual scaling with $N^{1/4}$ comes from. Due to the simplicity of the model we could trace back the problem to a central limit theorem for a sum of Bernoulli random variables \`a la Lyapunov~\cite{Lyapunov}. We think that in general this might be the correct approach since in the end one performs closed contour integrals which essentially ask how many eigenvalues of a complex matrix are encircled and with which winding.

In summary, we think that most of the statistics observed in class AIII may be super-universal meaning that they may hold for a large set of symmetry classes as well. This certainly has to be shown, though our results corroborate this conjecture.

\ack

This work was partly funded by the German–Israeli Foundation within the project Statistical Topology of Complex Quantum Systems, grant number I-1499-303.7/2019 (O.G. and T.G.) and partly by the Australian Research Council via the Discovery Project grant DP250102552 (M.K.).

\appendix

\section{Derivation of the k-Point Correlatior}
\label{sec:kpoint}

We first change the set of variables as follows
\begin{equation}
	p_n \rightarrow p_n + J_n , \quad q_n \rightarrow p_n.
\end{equation}
Based on this step we  now  expand the constituents of the generator up to first order in a source variable $J$, which will be set to zero after computing the derivative in Eq. \eqref{eq:ckconstruct}. To avoid working with poles, which appear not only in the determinant but also in the kernel function $\widehat K_2$, we cancel them by transforming the generator
\begin{align}
	\prod_{n = 1}^{k} \frac{\sin J_n}{\sin J_n} \, Z_{k}(p,p+J) = \frac{\displaystyle\mathrm{Pf}  \begin{bmatrix}
			\sin J_n\sin J_m\,\widehat{K}_{1}(p_m+J_m,p_n+J_n) &  \sin J_m\,\widehat K_{2}(p_m+J_m,p_n)  \\
			- \sin J_n\,\widehat K_{2}(p_n+J_n,p_m) & \widehat K_3(p_m,p_n)
		\end{bmatrix}_{1 \leq m,n \leq k } }{ \det \begin{bmatrix}
			\displaystyle \frac{\sin J_n}{\sin(p_m-p_n+J_m) }
		\end{bmatrix}_{1 \leq m,n \leq k}} .
\end{align}
For this purpose, we define the abbreviations
\begin{align}
	F_1 = & \det \begin{bmatrix} \displaystyle \frac{\sin J_n}{\sin(p_n - p_m + J_n)}
	\end{bmatrix}_{1 \leq n,m \leq k}^{-1},\\
	F_2 = & \mathrm{Pf}  \begin{bmatrix}
			\sin J_n\sin J_m\,\widehat{K}_{1}(p_m+J_m,p_n+J_n) &  \sin J_m\,\widehat K_{2}(p_m+J_m,p_n)  \\
			- \sin J_n\,\widehat K_{2}(p_n+J_n,p_m) & \widehat K_3(p_m,p_n)
		\end{bmatrix}_{1 \leq m,n \leq k } 
\end{align}
and apply the generalised Leibniz rule to the generator
\begin{align}
	\prod_{n = 1}^k\frac{\partial}{  \partial J_{n}} F_1 F_2 \Bigg \vert_{J =  0} = \sum \limits_{\sigma \in \mathbb{S}_{k}} \sum \limits_{m = 0}^{k} \frac{1}{m! (k-m)! } \left ( \prod \limits_{l = 1}^{m} \frac{\partial}{\partial J_{\sigma(l)}} \right ) F_1 \left ( \prod_{l = m +1}^{k} \frac{\partial}{\partial J_{\sigma(l)}} \right ) F_2 \Bigg \vert_{J = 0}
\end{align}
where the sum over the symmetric group $\mathbb{S}_{k}$ goes through all possible combinations. Due to overcounting we need to properly normalise by $m!$ and $(k-m)!$ as any permutation in the first and second product keeps them the same, respectively.

Firstly, we consider $F_1$ and expand its argument up to first order in each $J_l$,
\begin{align} 
	F_1  = \det \begin{bmatrix} \displaystyle 
		\delta_{nm} + (1-\delta_{nm}) \frac{J_n}{\sin(p_n-p_m)}
	+O(J_n^2)\end{bmatrix}_{1 \leq n,m \leq k}^{-1} \label{eq:f1approx},
\end{align}
This determinant can be expanded into a series of permanents using MacMahon's master theorem~\cite{MacMahon1915,Vitaly2022}. The idea is to write the determinant in terms of a complex Gaussian integral
\begin{equation}
F_1=\int \limits_{\mathbb{C}^k}\exp\left[\sum_{a,b=1}^kJ_a\frac{\delta_{ab}-1}{\sin(p_a-p_b)}v_a^*v_b\right]\prod_{j=1}^k\frac{e^{-|v_j|^2}d^2v_j}{\pi}.
\end{equation}
Then the derivative can be taken so that one is left with an integral that can be evaluated by the Wick-Isserlis theorem for complex random variables, i.e.,
\begin{equation}
\begin{split}
\left ( \prod \limits_{l = 1}^{m} \frac{\partial}{\partial J_{\sigma(l)}} \right ) F_1\Bigg \vert_{J = 0}=& \int \limits_{\mathbb{C}^k}\prod \limits_{l = 1}^{m} \left[\sum_{b=1}^k\frac{\delta_{\sigma(l)b}-1}{\sin(p_{\sigma(l)}-p_b)}v_{\sigma(l)}^*v_b\right]\prod_{j=1}^k\frac{e^{-|v_j|^2}d^2v_j}{\pi},\\
=&\sum_{\omega\in\mathbb{S}_m}\prod \limits_{l = 1}^{m} \frac{\delta_{\sigma(\omega(l))\sigma(l)}-1}{\sin(p_{\sigma(\omega(l))}-p_\sigma(l))}=\mathrm{perm}  \begin{bmatrix}
			\displaystyle\frac{\delta_{ab}-1}{\sin(p_{\sigma(a)} - p_{\sigma(b)})}
		\end{bmatrix}_{1 \leq a,b \leq m}.
\end{split}\label{eq:f1derivative}
\end{equation}
This permanent always vanishes when $m$ is odd since the matrix is totally antisymmetric. This allows to swap the sign in the permanent as it has been done in the result~\eqref{eq:kpointcorrelator}.

Next, we turn to compute the derivatives of  $F_2$. We express the argument in terms of diagonal and off-diagonal blocks. The antisymmetry of $\widehat K_1(p_n+J_n,p_m+J_m)$ and $\widehat K_3(p_n,p_m) $, see Eq. \eqref{eq:defd2}-\eqref{eq:defd4}, tells us that they must vanish for $n=m$. In contrast, $\widehat K_2(p_n+J_n,p_n)$ has a pole given by $J_{n}^{-1}$, which is compensated by the factor $J_n$ so that
\begin{equation}
\begin{split}
	F_2 =&  \mathrm{Pf} \left [   \delta_{nm}  \begin{bmatrix} 0 & 1+J_nC_1(p_n) \\
		-1-J_n C_1(p_n) & 0
	\end{bmatrix} + (1-\delta_{nm} )\begin{bmatrix}
			 J_nJ_m\widehat{K}_{1}(p_m,p_n) &  J_m\widehat K_{2}(p_m,p_n)  \\
			- J_n\widehat K_{2}(p_n,p_m) & \widehat K_3(p_m,p_n)
		\end{bmatrix}\right ]\\
		&\times\biggl[1+O\left(\sum_{l=1}^kJ_k^2\right)\biggl]
\end{split}\label{eq:f2approximation} 
\end{equation}
with $m,n=1,\ldots,k$ and the one-point correlator
\begin{equation}
\begin{split}
	C_1(p_n) =&  \frac{\partial}{\partial J_{n}} Z_{1|1}(p_n+J_n,p_n) \Bigg \vert_{J= 0}= \lim_{J_n\to0}\frac{\partial}{\partial J_n}J_n\widehat K_{2}(p_n+J_n,p_n)\\
	=&(-1)^{N}\tan (2p_n)  +(2N+1) \frac{\sin^{2N+1}(2p_n) }{\cos(2p_n)} q_{2N}^{(N+1)} \left ( \frac{1}{i \sin(2p_n)}\right ).
\end{split}
\end{equation}
The derivative in $J_{\sigma(m+1)},\ldots,J_{\sigma(k)}$ at $J=0$ selects the corresponding rows and columns. Thus, we arrive at
\begin{equation}
\begin{split}
	&\prod_{l = m + 1}^{k} \frac{\partial }{\partial J_{\sigma(l)}} F_2\Bigl\vert_{J=0}\\
	 =&  \mathrm{Pf} \left [  \delta_{ab}\begin{bmatrix}
		0 & C_{1}(p_{\sigma(a)}) \\
		- C_{1}(p_{\sigma(a)}) & 0 
	\end{bmatrix}  + (1- \delta_{ab}) \begin{bmatrix}
			\widehat{K}_{1}(p_{\sigma(a)},p_{\sigma(b)}) &  \widehat K_{2}(p_{\sigma(a)},p_{\sigma(b)})  \\
			- \widehat K_{2}(p_{\sigma(b)},p_{\sigma(a)}) & \widehat K_3(p_{\sigma(a)},p_{\sigma(b)})
		\end{bmatrix}\right ].
\end{split}\label{eq:f2derivative}  
\end{equation}
with $m + 1\leq a,b \leq k$. Once we put everything together we obtain the result~\eqref{eq:kpointcorrelator}.

\section{Derivations of the terms $D_2$, $D_3$ and $D_4$}\label{sec:Derivations.D}

\subsection{ Kernel $D_3$}
\label{secd3}

We start with the term $D_3$ instead of $D_2$ as we will need an intermediate result of $D_3$ to simplify the $D_2$ term.

The integrand for $D_3$ is free of any singularities as it is only a polynomial in $\sin^2\Delta$ and $\cos^2\Omega$, see~\eqref{eq:defd3}. In particular, it is evident that the integrand has a periodicity of $\pi$ and not $2\pi$ in the variable $\Delta$. Thus, we can go over to a complex variable $z=e^{2i\Delta}$ and use the representation~\eqref{qhyp} of the polynomials $q_{2(N-\alpha)}^{(N+1-\alpha)}(x)$ in terms of the hypergeometric function so that
\begin{equation}
\begin{split}
	D_3 = & -\frac{N(2N+1)}{2} \int \limits_0^{4\pi} {\rm d}\Omega \oint \limits_{|z|=1} \frac{{\rm d}z}{i z^2} (z-1)^2  \cos^{4N-2}\Omega\\
	&\times  \hyp{-N }{1}{3/2}{-\frac{(z-1)^2}{4z\cos^2\Omega}}\hyp{1-N }{1}{3/2}{-\frac{(z-1)^2}{4z\cos^2\Omega}}.
\end{split}
\end{equation}
As the origin and infinity are the only singularities on the Riemann sphere for $z$ we can choose an arbitrary radius $r$ for its contour. We choose $0<r<1$.

Next, we employ a Pfaff transformation of the hypergeometric function, especially
\begin{equation}
\hyp{\alpha-N }{1}{3/2}{x}=(1-x)^{-1}\hyp{N+3/2-\alpha }{1}{3/2}{\frac{x}{x-1}}
\end{equation}
for both polynomials so that
\begin{equation}
\begin{split}
	D_3 = &8 iN(2N+1) \int \limits_0^{4\pi} {\rm d}\Omega \oint \limits_{|z|=r} {\rm d}z   \frac{(z-1)^2\cos^{4N+2}\Omega}{[4z\cos^2\Omega+(z-1)^2]^2}\\
	&\times  \hyp{N+3/2 }{1}{3/2}{\frac{(z-1)^2}{(z-1)^2+4z\cos^2\Omega}}\hyp{N+1/2 }{1}{3/2}{\frac{(z-1)^2}{(z-1)^2+4z\cos^2\Omega}}.
\end{split}
\end{equation}
These hypergeometric functions have the explicit forms
\begin{equation}
\hyp{N+3/2 }{1}{3/2}{\zeta}=\frac{\Gamma[3/2]}{\Gamma[N+3/2]}\partial_x^{N}\frac{x^{N+1/2}}{1-x\zeta}\Biggl|_{x=1} 
\end{equation}
and
\begin{equation}
 \hyp{N+1/2 }{1}{3/2}{\zeta}=\frac{\Gamma[3/2]}{\Gamma[N+1/2]}\partial_y^{N-1}\frac{y^{N-1/2}}{1-y\zeta}\Biggl|_{y=1}.
\end{equation}
Hence, it is
\begin{equation}
\begin{split}
	D_3 = &\frac{4\pi iN}{\Gamma^2[N+1/2]} \int \limits_0^{4\pi} {\rm d}\Omega \cos^{4N+2}\Omega\oint \limits_{|z|=r} {\rm d}z   \partial_x^{N}\partial_y^{N-1}\frac{x^{N+1/2}y^{N-1/2}}{x-y}\\
	&\times  \left[\frac{1}{(1-x)(z-1)^2+4z\cos^2\Omega}-\frac{1}{(1-y)(z-1)^2+4z\cos^2\Omega}\right]_{x,y=1}.
\end{split}
\end{equation}
The derivatives and their limits can be exchanged with the contour integral for a fixed $\Omega$ with $\cos\Omega\neq0$. We recall that, once we carry out the contour integral and derivative, the integrand is a polynomial in $\cos^2\Omega$ so that the points $\cos\Omega=0$ are of measure zero.

It can be readily checked that one pole in $z$ of each of the two terms is either very close to the origin or to infinity when $x,y\to1$ and $\cos\Omega\neq0$. 
 Therefore, we can carry out the contour integral via the residue theorem where the contributing poles are
\begin{equation}
z=1-\frac{2\cos^2\Omega}{1-x}\left[1-\sqrt{1-\frac{1-x}{\cos^2\Omega}}\right] \quad{\rm and}\quad z=1-\frac{2\cos^2\Omega}{1-y}\left[1-\sqrt{1-\frac{1-y}{\cos^2\Omega}}\right]
\end{equation}
such that
\begin{equation}
\begin{split}
	D_3 = &\frac{-2\pi^2 iN}{\Gamma^2[N+1/2]} \int \limits_0^{4\pi} {\rm d}\Omega \cos^{4N}\Omega  \partial_x^{N}\partial_y^{N-1}\frac{x^{N+1/2}y^{N-1/2}}{x-y}\\
	&\times\left[\frac{1}{\sqrt{1-(1-x)/\cos^2\Omega}}-\frac{1}{\sqrt{1-(1-y)/\cos^2\Omega}}\right]_{x,y=1}.
\end{split}
\end{equation}
This expression allows us to expand in the square roots as we send $x,y\to1$ and before we integrate over $\Omega$ such that we can assume $\cos\Omega\neq0$ which leaves us with
\begin{equation}
\begin{split}
	D_3 = &\frac{-4\pi^2 iN}{\Gamma^2[N+1/2]} \sum_{j=1}^{2N}\binom{-1/2}{j}\binom{4N-2j}{2N-j}\frac{1}{2^{4N-2j}}  \partial_x^{N}\partial_y^{N-1}x^{N+1/2}y^{N-1/2}\frac{(x-1)^j-(y-1)^j}{x-y}\Biggl|_{x,y=1}\\
	= &\frac{-4\pi^2 iN}{2^{4N}\Gamma^2[N+1/2]} \binom{4N}{2N}\partial_x^{N}\partial_y^{N-1}(1-x)^{N+1/2}(1-y)^{N-1/2}\\
	&\times\frac{\hyp{-2N }{1/2}{1/2-2N}{x}-\hyp{-2N }{1/2}{1/2-2N}{y}}{x-y}\Biggl|_{x,y=0},
\end{split}
\end{equation}
where we have shifted in the second line $x\to1-x$ and $y\to1-y$. The second expression can be obtained when writing
\begin{equation}
\frac{\hyp{-2N }{1/2}{1/2-2N}{x}-\hyp{-2N }{1/2}{1/2-2N}{y}}{x-y}=-\int \limits_0^1 { _{2}F'_1}\left(\left.\begin{array}{c} -2N,\, 1/2\\ 1/2-2N\end{array}\right|x+(y-x)t\right){\rm d}t,
\end{equation}
using the series representation of the hypergeometric function which terminates into a sum and expanding the binomials $(1-x)^{N+1/2}(1-y)^{N-1/2}$.

When carrying out the derivatives and using the sum representation of the hypergeometric function, the term $D_3$ is proportional to the double sum
\begin{equation}\label{D3.prelim}
D_3\propto \sum_{a=0}^N\sum_{b=0}^{N-1}(-1)^{a+b}\binom{N+1/2}{a}\binom{N-1/2}{b}B\left(a+b+\frac{1}{2},\frac{1}{2}\right)B\left(2N-a-b+\frac{1}{2},\frac{1}{2}\right).
\end{equation}
To show\footnote{The main ideas of the following calculation, especially the proof to show $D_3=0$, though not the full details have been obtained with the help of ChatGPT 5.6. Sol} that this double sum vanishes exactly we consider the more general sum
\begin{equation}
S_{l,k}=\sum_{a=0}^l\sum_{b=0}^{k}(-1)^{a+b}\binom{l+1/2}{a}\binom{k+1/2}{b}B\left(a+b+\frac{1}{2},\frac{1}{2}\right)B\left(l-a+k-b+\frac{3}{2},\frac{1}{2}\right)=S_{k,l}
\end{equation}
where we are specifically interested in $S_{N,N-1}$. To find the symmetry which tells us that $S_{N,N-1}$ vanishes, we consider the generating function
\begin{equation}
F(x,y)=\sum_{l,k=0}^\infty S_{l,k}x^ly^k
\end{equation}
with the auxiliary variables $x,y<1$ so that the series converges absolutely.

In the first step we rewrite the two Beta functions in terms of integrals
\begin{equation}
B\left(a+b+\frac{1}{2},\frac{1}{2}\right)B\left(l-a+k-b+\frac{3}{2},\frac{1}{2}\right)=\int \limits_0^1ds\int \limits_0^1\frac{s^{a+b}t^{l-a+k-b+1}{\rm d}s{\rm d}t}{\sqrt{s(1-s)}\sqrt{t(1-t)}}.
\end{equation}
The fourfold sums are then factorising when interchanged with the integrals
\begin{equation}
F(x,y)=\int \limits_0^1\int \limits_0^1 H(xt,xs)H(yt,ys)\frac{t\,{\rm d}s{\rm d}t}{\sqrt{s(1-s)}\sqrt{t(1-t)}}
\end{equation}
depending on the function
\begin{equation}
H(xt,xs)=\sum_{l=0}^\infty \sum_{a=0}^l(-1)^a\binom{l+1/2}{a}(xt)^l\left(\frac{s}{t}\right)^a=\sum_{a=0}^\infty\sum_{l=0}^\infty \binom{l+a+1/2}{a}(xt)^l(-xs)^a,
\end{equation}
where in the second equality we have interchanged the series and shifted $l\to l+a$. The sum over $a$ is simply a binomial series
\begin{equation}
\sum_{a=0}^\infty\binom{l+a+1/2}{a}(-xs)^a=\frac{1}{(1+xs)^{l+3/2}},
\end{equation}
so that the sum over $l$ becomes a geometric series
\begin{equation}
H(xt,xs)=\sum_{l=0}^\infty \frac{(xt)^l}{(1+xs)^{l+3/2}}=\frac{1}{(1+xs-xt)\sqrt{1+xs}}.
\end{equation}
This means that we need to consider the integral
\begin{equation}\label{Fxy.a}
F(x,y)=\int \limits_0^1\int \limits_0^1 \frac{t\,{\rm d}s{\rm d}t}{(1+xs-xt)(1+ys-yt)\sqrt{s(1-s)(1+xs)(1+ys)}\sqrt{t(1-t)}}.
\end{equation}
Next we integrate over $t$ by using the partial fraction decomposition
\begin{equation}
\frac{t}{(1+xs-xt)(1+ys-yt)}=\frac{1}{x-y}\left[\frac{1+xs}{1+xs-xt}-\frac{1+ys}{1+ys-yt}\right]
\end{equation}
and the integral identity
\begin{equation}
\int \limits_0^1\frac{{\rm d}t}{(1+xs-xt)\sqrt{t(1-t)}}=\frac{\pi}{\sqrt{(1+xs)(1-x[1-s])}}
\end{equation}
and similarly for $x$ replaced by $y$, which is the Cauchy transform of the arcsine law. Plugging this back into~\eqref{Fxy.a} we arrive at
\begin{equation}\label{Fxy.b}
F(x,y)=\int \limits_0^1\frac{\pi{\rm d}s}{(x-y)\sqrt{s(1-s)}}\left[\frac{1}{\sqrt{(1+ys)(1-x[1-s])}}-\frac{1}{\sqrt{(1+xs)(1-y[1-s])}}\right].
\end{equation}
This expression shows that the function $F(x,y)$ satisfies the symmetry $F(x,y)=F(-y,-x)$ which becomes apparent when substituting $s\leftrightarrow 1-s$. This however implies for the coefficients
\begin{equation}
\sum_{l,k=0}^\infty S_{l,k}x^ly^k=F(x,y)=F(-y,-x)=\sum_{l,k=0}^\infty (-1)^{l+k}S_{k,l}x^ly^k
\end{equation}
or simply $ S_{l,k}=(-1)^{l+k}S_{k,l}$. Together with the symmetry $S_{l,k}=S_{k,l}$ this implies $S_{N,N-1}=0$ or equivalently $D_3=0$.

\subsection{Kernel $D_2$} 
\label{secd2}

In the first step to compute $D_2$, see~\eqref{eq:defd2}, we make use of the sum representation~\eqref{qhyp} of the polynomials $q_{2N}^{(N+1)}(x)$ and combine it with the trigonometric identity
\begin{equation}
	 \cos(\Omega + \Delta)\cos(\Omega-\Delta)= \cos^{2} \Delta -\sin^{2} \Omega .
\end{equation}
In particular, we focus on the term
\begin{equation}
\begin{split}
&\frac{\sin^{4N+2} \Omega \ q_{2N}^{(N+1)}\left( - i \cos \Delta / \sin \Omega\right)^{2}}{\cos(\Omega-\Delta) \cos( \Omega +\Delta) }\\
=&\sum_{m,n=0}^{N}\frac{(-1)^{m+n}B(N+1,3/2)^2}{B(m+1,N-m+3/2)B(n+1,N-n+3/2)}\frac{\sin^{4N+2-2m-2n} \Omega \ \cos^{2m+2n} \Delta}{\cos(\Omega-\Delta) \cos( \Omega +\Delta) },
\end{split}
\end{equation}
which can be simplified by
\begin{equation}
	\frac{\sin^{4N+2-2m-2n }\Omega \cos^{2m+2n} \Delta}{\cos(\Omega-\Delta) \cos(\Omega+\Delta)} = \frac{\cos^{4N+2} \Delta}{\cos(\Omega-\Delta) \cos(\Omega+\Delta)} -\sum_{j = m+n}^{2N}  \cos^{2j} \Delta \sin^{4N-2j} \Omega. \label{rec}
\end{equation}
Then it is
\begin{equation}
\begin{split}
&\frac{\sin^{4N+2} \Omega \ q_{2N}^{(N+1)}\left( - i \cos \Delta / \sin \Omega\right)^{2}}{\cos(\Omega-\Delta) \cos( \Omega +\Delta) }=\frac{\cos^{4N+2} \Delta  }{\cos(\Omega-\Delta) \cos(\Omega+\Delta)}q_{2N}^{(N+1)}\left( - i\right)^{2}\\
-&\sum_{m,n=0}^{N}\sum_{j = m+n}^{2N} \frac{(-1)^{m+n}B(N+1,3/2)^2}{B(m+1,N-m+3/2)B(n+1,N-n+3/2)}\cos^{2j} \Delta \sin^{4N-2j} \Omega.
\end{split}
\end{equation}
Using the normalisation of the polynomials $q_{2N}^{(N+1)}\left( - i\right)^{2}=1/(2N+1)^2$, it becomes clear that the first of this expression cancels with the first term of the integrand of $D_2$ in~\eqref{eq:defd2} so that
\begin{equation}
\begin{split}
	D_2 =&  \sum_{n,m = 0}^{N} \sum_{j = m + n}^{2N} \frac{(-1)^{m+n}B(N+1,3/2)^{2} }{B(m+1,N-m+3/2)B(n+1,N-n+3/2)}\int \limits_0^{4\pi}{\rm d}\Omega \sin^{4N-2j} \Omega \intl \Delta \cos^{2j} \Delta \\
	=&8(2N+1)^{2} \sum_{n,m = 0}^{N} \sum_{j = n + m}^{2N} \frac{(-1)^{n+m} B(N+1,3/2)^{2} B(1/2,j+1/2) B(1/2,2N-j+1/2) }{B(n+1,N-n+3/2)B(m+1,N-m+3/2)}
\end{split} 
\end{equation}
To simplify this expression we make use of the fact that $D_3$ is vanishing, especially that the sum~\eqref{D3.prelim} vanishes\footnote{The main ideas of the ensuing simplification though not the full details have been gained from ChatGPT 5.6. Sol}.

For this purpose we first rewrite
\begin{equation}
\frac{1 }{B(n+1,N-n+3/2)B(m+1,N-m+3/2)}=\frac{(2N+3)^2}{4}\binom{N+1/2}{n}\binom{N+1/2}{m}
\end{equation}
and interchange the sum over $j$ and with the double sum over $n$ and $m$, i.e.,
\begin{equation}
\sum_{n,m = 0}^{N} \sum_{j = m + n}^{2N}=\sum_{j=0}^{2N}\sum_{\substack{0\leq n,m\leq N\\n+m\leq j}}.
\end{equation}
Then the sum over $n$ and $m$ can be understood as a contour integral,
\begin{equation}\label{der.D2.a}
\sum_{\substack{0\leq n,m\leq N\\n+m\leq j}}(-1)^{n+m}\binom{N+1/2}{n}\binom{N+1/2}{m}=\oint \limits_{|z|=1/2}\frac{{\rm d}z}{2\pi i z^{j+1}(1-z)}\left(\sum_{a=0}^N(-z)^a\binom{N+1/2}{a}\right)^2.
\end{equation}
We denote the sum by 
\begin{equation}
A_N(z)=\sum_{a=0}^N(-z)^a\binom{N+1/2}{a}.
\end{equation}
Then, direct calculation gives  the following differential equation
\begin{equation}
(1-z)\partial_zA_N(z)=-\frac{2N+1}{2}A_N(z)+\frac{(-z)^N}{B(1/2,N+1)}.
\end{equation}
We replace $A_N(z)/(1-z)$ accordingly in~\eqref{der.D2.a} to find for the double sum
\begin{equation}
\begin{split}
&\sum_{\substack{0\leq n,m\leq N\\n+m\leq j}}(-1)^{n+m}\binom{N+1/2}{n}\binom{N+1/2}{m}\\
=&\oint \limits_{|z|=1/2}\frac{{\rm d}z}{2\pi i z^{j+1}}\left[\binom{N-1/2}{N}\frac{(-z)^N}{1-z}A_N(z)-\frac{2A_N(z)\partial_zA_N(z)}{2N+1}\right].
\end{split}
\end{equation}
The integral of the second term is actually
\begin{equation}
\oint \limits_{|z|=1/2}\frac{{\rm d}z}{2\pi i z^{j+1}}\frac{2A_N(z)\partial_zA_N(z)}{2N+1}=\sum_{n=0}^N\sum_{m=0}^{N-1}(-1)^{n+m}\binom{N+1/2}{n}\binom{N-1/2}{m}\delta_{n+m,j},
\end{equation}
where $\delta_{m,n}$ is the Kronecker symbol. When summing over $j=0,\ldots, 2N$ with the Beta functions $B(1/2,j+1/2) B(1/2,2N-j+1/2)$ we obtain the sum~\eqref{D3.prelim} for which we know that it vanishes.

Hence, the term $D_2$ is equal to
\begin{equation}
\begin{split}
	D_2 =&2(2N+1)^{2}(2N+3)^{2}B^{2}\left(N+1,\frac{3}{2}\right) \binom{N-1/2}{N}\sum_{j = 0}^{2N} B\left(\frac{1}{2},j+\frac{1}{2}\right) B\left(\frac{1}{2},2N-j+\frac{1}{2}\right) \\
	&\times\oint \limits_{|z|=1/2}\frac{{\rm d}z}{2\pi i z^{j+1}}\frac{(-z)^N}{1-z}A_N(z)\\
	=&(-1)^N8\binom{N-1/2}{N}^{-1}\sum_{j = N}^{2N} \sum_{a=0}^{j-N}(-1)^aB\left(\frac{1}{2},j+\frac{1}{2}\right) B\left(\frac{1}{2},2N-j+\frac{1}{2}\right)\binom{N+1/2}{a}.
\end{split}
\end{equation}
The sum over $a$ is an alternating binomial sum which gives
\begin{equation}
\sum_{a=0}^{j-N}(-1)^a\binom{N+1/2}{a}=(-1)^{j-N}\binom{N-1/2}{j-N}.
\end{equation}
When shifting $j\to j+N$ we arrive at
\begin{equation}
\begin{split}
	D_2 =&(-1)^N8\binom{N-1/2}{N}^{-1}\sum_{j = 0}^{N}(-1)^jB\left(\frac{1}{2},N+j+\frac{1}{2}\right) B\left(\frac{1}{2},N-j+\frac{1}{2}\right)\binom{N-1/2}{j}\\
	=&(-1)^N8\pi\sum_{j = 0}^{N}(-1)^jB\left(\frac{1}{2},N+j+\frac{1}{2}\right)\binom{N}{j}.
\end{split}
\end{equation}
The Beta function can be replaced by its integral representations so that the binomial sum can be carried out,
\begin{equation}
\begin{split}\label{d2singularregular}
	D_2 =&(-1)^N8\pi\sum_{j = 0}^{N}(-1)^j\int \limits_0^1\frac{t^{N+j}{\rm d}t}{\sqrt{t(1-t)}}\binom{N}{j}=(-1)^N8\int \limits_0^1t^{N-1/2}(1-t)^{N-1/2}{\rm d}t\\
	=&(-1)^N8\pi B(N+1/2,N+1/2).
\end{split}
\end{equation}

\subsection{Kernel $D_4$}
\label{secd4}

When calculating $D_4$, it is signifcantly more convenient to integrate the $\Delta$-dependency in Eq.~\eqref{eq:d4ersterschritt} out first.
\begin{equation}
\begin{split}
	D_4 = &  (-1)^{N}2N (2N+1) \sum_{m=0}^{N-1}(-1)^m\frac{B(N,3/2)}{B(m+1,N-m+1/2)}  \\
	&\times \int \limits_0^{4\pi} {\rm d}\Omega \int \limits_0^{2\pi}  {\rm d}\Delta \Phi_{2N+2}^{(1)} \left ( \cos^{2} \Omega \right )\cos^{2N+2+2m} \Omega\  \sin^{2N-2m} \Delta \label{eq:d4ersterschritt} 
\end{split}
\end{equation}
The integral over $\Delta$ contributes essentially a beta function. $2B(1/2,N-m+1/2)$, which cancels with some terms in the denominator leading to
\begin{equation}
\begin{split}
	D_4 
	=&(-1)^{N+1}2\pi(2N+1) \sum_{m=0}^{N-1}(-1)^m\binom{N}{m}\int \limits_0^{4\pi} {\rm d}\Omega \Phi_{2N+2}^{(1)} \left ( \cos^{2} \Omega \right)\cos^{2N+2+2m} \Omega.
\end{split}\label{eq:d4zwischenschritt}
\end{equation}
This expression invites to perform the truncated binomial sum which simplifies the integral drastically,
\begin{equation}
	D_4 =(-1)^{N}2\pi(2N+1) \int \limits_0^{4\pi} {\rm d}\Omega \Phi_{2N+2}^{(1)} \left ( \cos^{2} \Omega \right)\cos^{2N+2} \Omega[\sin^{2N}\Omega-(-1)^N\cos^{2N}\Omega]
\end{equation}
Before we proceed further we split up  the integration domain into eight intervals where the map $ \omega = \cos^{2} \Omega $ is bijective. They all contribute the same due to symmetry so that
\begin{equation}
	D_4 = (-1)^{N}8\pi (2N+1)  \int \limits_{0}^{1}  \mathrm{d} \omega \left  [ \omega^{N+1/2} (1-\omega)^{N-1/2} - (-1)^N\omega^{2N+1/2} (1-\omega)^{-1/2} \right ] \Phi_{2N+2}^{(1)} ( \omega ) \label{wintegral}.
\end{equation}
When using the series representation of Lerch's transcendental function~\eqref{eq:lerch} we find
\begin{equation}
\begin{split}
\int \limits_{0}^{1}\mathrm{d} \omega \omega^{N+1/2} (1-\omega)^{N-1/2}\Phi_{2N+2}^{(1)} ( \omega )=&\sum_{j=0}^\infty\frac{\Gamma[N+j+3/2]\Gamma[N+1/2]}{\Gamma[2N+j+2](2N+j+2)}=\frac{2}{2N+1}B(N+3/2,N+1/2),\\
\int \limits_{0}^{1}\mathrm{d} \omega \omega^{2N+1/2} (1-\omega)^{-1/2}\Phi_{2N+2}^{(1)} ( \omega )=&\sum_{j=0}^\infty\frac{\Gamma[2N+j+3/2]\Gamma[1/2]}{\Gamma[2N+j+2](2N+j+2)}=2B(2N+3/2,1/2).
\end{split}
\end{equation}
Plugging this into~\eqref{wintegral} leads to
\begin{equation}
	D_4 = 8\pi   \left[(-1)^{N}B(N+1/2,N+1/2)+B(2N+3/2,-1/2)\right].
\end{equation}

\section{Calculation of the probability $r(m)$}
\label{sec:rm}

We start from~\eqref{eq: r integrals} and go over to polar coordinates $ z_n = r_n e^{i \phi_n} $. Once the Vandermonde determinant $\Delta_{2N}(\omega)$ is written in terms of Leibniz formula the integrals factorise, i.e.,
\begin{align}
	r(m) = \frac{1}{c^{(4)}}\sum_{ \sigma \in \mathbb{S}_{2N} } \mathrm{sgn} \,\sigma  \, & \left ( \prod_{n = 1}^{m} \intla{0}{1} r_n\frac{r_{n}^{\sigma(2n)+ \sigma(2n-1)}}{(1+r_{n}^{2})^{2N+2}}\right ) \left (\prod_{n = m +1}^{N} \intla{1}{\infty} r_{n} \frac{r_{n}^{\sigma(2n)+ \sigma(2n-1)}}{(1+r_{n}^{2})^{2N+2}} \right ) \nonumber \\
	& \times\prod_{n = 1}^{N} \intl \phi_{n} \left [e^{i \phi_n (\sigma(2n-1) - \sigma(2n) +1 )}   -  e^{i \phi_n ( \sigma(2n-1) -\sigma(2n) - 1) }\right ].
\end{align}
For the angular part we obtain a linear combination of Kronecker deltas
\begin{align}
	\intl \phi_{n} \left [e^{i \phi_n (\sigma(2n-1) - \sigma(2n) +1 )}   -  e^{i \phi_n ( \sigma(2n-1) -\sigma(2n) - 1) }\right ] = & 2 \pi  [\delta_{\sigma(2n-1)+1,\sigma(2n) } - 2  \delta_{\sigma(2n-1)+1,\sigma(2n)}] . \label{eq:negativeangularpart}
\end{align}
Both Kronecker deltas restrict the possible permutation elements in $\sigma$ to be paired in tuples of odd and even integers, that only differ by a unit step. Thus the summation over $\mathbb S_{2N}$ is reduced to $\mathbb S_N$ with a new permutation identified as follows $(\sigma(2n),\sigma(2n-1))\mapsto(2\sigma'(n),2\sigma'(n)-1)$. The radial integrals effectively become
\begin{align}
	\intla{0}{1} r_n\frac{r_{n}^{\sigma(2n)+ \sigma(2n-1)}}{(1+r_{n}^{2})^{2N+2}} = \intla{0}{1} r_n\frac{r_{n}^{4\sigma'(n)-1}}{(1+r_{n}^{2})^{2N+2}}=\frac{1}{2}B(2\sigma'(n),2N-2\sigma'(n)+2)p_{\sigma'(n)}
\end{align}
with the incomplete beta functions~\eqref{incomplete.Beta} as in \cite{Braun2022}. Due to the negative sign in the second term of \eqref{eq:negativeangularpart} the signum  cancels out when permuting inside a pair $(2j,2j-1)$ while permuting between pairs consists of always of an even permutation. The product of the beta functions  is also cancelled by the normalisation constant $c^{(4)}$ and we arrive at the final form~\eqref{eq: r(m)}.

\section{Asymptotic of key expression in the correlators }
\label{sec:unfold.der}

The asymptotic of the skew-orthogonal polynomials for $q_{2(N-\alpha)}^{(N+1-\alpha)}(x)$ for fixed $\alpha$ but double scaling limits $\sqrt{N}x\to\widetilde{x}$ and $x/\sqrt{N}\to\widetilde{x}$ when $N\to\infty$ is easiest found by the sum representation~\eqref{qhyp}
and the behaviour
\begin{equation}
B(a,N+b)\overset{N\to\infty}{\sim} \Gamma[a]N^{-a}
\end{equation}
for any fixed $a$ and $b$. The asymptotic is then
\begin{equation}\label{q.lim.a}
q_{2(N-\alpha)}^{(N+1-\alpha)}(x)\overset{N\to\infty}{\sim}\left\{\begin{array}{cl} \displaystyle \sqrt{\frac{\pi}{4N}}\exp[\widetilde{x}^2], & \sqrt{N}x\to\widetilde{x},\\ \displaystyle \sqrt{\frac{\pi}{4}}(\sqrt{N}\widetilde{x})^{2(N-\alpha)}\exp[\widetilde{x}^{-2}]\frac{{\rm erfi}[i\widetilde{x}^{-1}]}{i\widetilde{x}^{-1}}, & x/\sqrt{N}\to\widetilde{x}, \end{array}\right.
\end{equation}
with the imaginary error function ${\rm erfi}(y)$. For $x$ being fixed in the large $N$ limit, we will use its representation as a hypergeometric function~\eqref{qhyp} and Euler's integral formula~\cite{DLMF}
\begin{align}
	x^{2(\alpha-N)}q_{2(N-\alpha)}^{(N+1-\alpha)}(x) =
\frac{1}{2}\int \limits_0^1\left(1+\frac{ t}{x^2}\right)^{N-\alpha}\frac{{\rm d}t}{\sqrt{1-t}}.
\end{align}
The maximum of the modulus exponential term in $N$ can be obtained only at the boundaries. We need to take into account that in the present case we have always $x^2<0$. Thus, the maximum is attained at $t=0$ when $x^2<-1/2$ and it is attained at $t=1$ when $x^2>-1/2$. For $x^2=-1/2$ both boundaries would contribute. However, the one at $t=0$ is algebraically suppressed to the contribution resulting from $t=1$. Summarising, it is
\begin{equation}\label{q.lim.b}
\begin{split}
x^{2(\alpha-N)}q_{2(N-\alpha)}^{(N+1-\alpha)}(x)\overset{N\to\infty}{\sim} \left\{\begin{array}{cl} \displaystyle \frac{|x^2|}{2N} ,& x^2<-1/2,\\ \displaystyle \sqrt{\frac{\pi(1+x^2)}{4N}}\left(1+\frac{1}{x^2}\right)^{N-\alpha},& x^2\geq-1/2. \end{array}\right.
\end{split}
\end{equation}
for fixed real $x^2$.

The other special function is the Lerch transcendent~\eqref{eq:lerch} which is however simpler and has only two asymptotic for an argument $x\in[0,1]$ for large $N$,
\begin{equation}\label{phi.lim}
\Phi_{2N+2}^{(1)} (x) \overset{N\to\infty}{\sim}\left\{\begin{array}{cl}\displaystyle \frac{1}{2N}\frac{1}{1-x},& 0\leq x<1,\\ \displaystyle \Gamma[0,\widetilde x]e^{\widetilde{x}}, &\displaystyle x=1-\frac{\widetilde{x}}{N}. \end{array}\right.
\end{equation}
with the incomplete gamma function $\Gamma[0,\widetilde{x}]$.
 The second branch follows by replacing the series~\eqref{eq:lerch} by a Riemann integral.

\newpage
\bibliographystyle{iopart-num}
\bibliography{ciiwindingnumber}
\end{document}